\documentclass[twocolumn,structabstract]{aa}
\usepackage{amsmath}
\usepackage{fixltx2e}
\usepackage[english]{babel}
\usepackage{graphicx}
\usepackage{epstopdf}
\usepackage{epsf,color}
\usepackage[mathscr]{eucal}
\usepackage{amsmath}
\usepackage{amssymb,amsfonts}
\usepackage{natbib}
\usepackage{graphicx}
\usepackage{txfonts}
\usepackage{dsfont}
\definecolor{Mygreen}{rgb}{0.75, 0.0, 0.0}
\definecolor{Mypink}{rgb}{1.0, 0.0, 0.5}
\definecolor{Myred}{rgb}{0.7, 0.0, 0.0}
\usepackage[breaklinks, citecolor=blue, linkcolor=Myred, urlcolor=Myred, colorlinks=true, debug, baseurl=' ']{hyperref}
\usepackage{float}
\usepackage{color}
\usepackage{ulem}
\usepackage{subcaption}
\usepackage{wasysym}

\bibpunct{(}{)}{;}{a}{}{,}
\begin{document}
\title{Reddening map and recent star formation {\bf in the} Magellanic Clouds based on OGLE IV Cepheids}
\author{
Y.~C.~Joshi$^{1}$\thanks{E-mail: yogesh@aries.res.in}\thanks{Data used in the present study is only available in the electronic form at the CDS via anonymous ftp to cdsarc.u-strasbg.fr},
A. Panchal$^{1}$
}
\institute{
$^{1}$Aryabhatta Research Institute of Observational Sciences (ARIES), Manora peak, Nainital 263002, India\\
}
\date{Received: 05 November 2018; accepted 11 June 2019}

\abstract
{The reddening maps of the Large Magellanic Cloud (LMC) and Small Magellanic Cloud (SMC) are constructed using the Cepheid Period-Luminosity ($P$-$L$) relations.}
{We examine reddening distribution across the LMC and SMC through largest data on Classical Cepheids provided by the OGLE Phase IV survey. We also investigate the age and spatio-temporal distributions of Cepheids to understand the recent star formation history in the LMC and SMC.}
{The $V$ and $I$ band photometric data of 2476 fundamental mode (FU) and 1775 first overtone mode (FO) Cepheids in the LMC and 2753 FU and 1793 FO Cepheids in the SMC are analyzed for their $P$-$L$ relations. We convert period of FO Cepheids to corresponding period of FU Cepheids before combining the two modes of Cepheids. Both galaxies are divided into small segments and combined FU and FO $P$-$L$ diagrams are drawn in two bands for each segment. The reddening analysis is performed on 133 segments covering a total area of about 154.6 deg$^2$ in the LMC and 136 segments covering a total area of about 31.3 deg$^2$ in the SMC. By comparing with well calibrated $P$-$L$ relations of these two galaxies, we determine reddening $E(V-I)$ in each segment and equivalent reddening $E(B-V)$ assuming the normal extinction law. The period-age relations are used to derive the age of the Cepheids.}
{Using reddening values in different segments across the LMC and SMC, reddening maps are constructed. We find clumpy structures in the reddening distributions of the LMC and SMC. From the reddening map of the LMC, highest reddening of $E(V-I) = 0.466$ mag is traced in the region centered at $\alpha \sim 85^{o}.13,~\delta \sim -69^{o}.34$ which is in close vicinity of the star forming HII region 30 Doradus. In the SMC, maximum reddening of $E(V-I) = 0.189$ mag is detected in the region centered at $\alpha \sim 12^{o}.10,~\delta \sim -73^{o}.07$. The mean reddening values in the LMC and SMC are estimated as $E(V-I)_{LMC} = 0.113\pm0.060$ mag, $E(B-V)_{LMC} = 0.091\pm0.050$ mag, $E(V-I)_{SMC} = 0.049\pm0.070$ mag, and $E(B-V)_{SMC} = 0.038\pm0.053$ mag.}
{The LMC reddening map displays heterogeneous distribution having small reddening in the central region and higher reddening towards eastern side of the LMC bar. The SMC have relatively small reddening in peripheral regions but larger reddening towards south-west region. In these galaxies, we see an evidence of a common enhanced Cepheid population at around 200 Myr ago which appears to have occurred due to close encounter between the two clouds.}
\authorrunning{Y. C. Joshi et al.}

\titlerunning{Reddening map and recent star formation in the Magellanic Clouds}

\maketitle
\section{Introduction}\label{intro}
The LMC and SMC (together known as Magellanic Clouds, MCs) are among one of the most studied galaxies in the Universe due to their close proximity to the Galaxy as they are located at the distance of $\approx$ 50 kpc and 60 kpc, respectively \citep{1991IAUS..148...15W, 2005MNRAS.357..304H, 2006ApJ...642..834K, 2015AJ....149..179D}. They offer an excellent opportunity to strive many astronomical issues such as star formation, structure formation, distribution of interstellar medium through wide range of tracers and they have significantly advanced our understanding on galaxy evolution and dynamical interaction \citep[e.g.,][]{2004AJ....127.1531H, 2014ApJ...795..108S, 2014MNRAS.442.1897R, 2015MNRAS.450..552P, 2017AJ....154..199N}. The MCs provide the ideal laboratory to probe the spatially resolved star formation history (SFH) and dust distributions since these galaxies are not just close enough to be resolved into stars but also moderately affected by the interstellar extinction and foreground Milky Way stars \citep{2018MNRAS.478.5017R}. In recent times, the distribution of different stellar populations in the MCs has been studied by various authors  \citep[e.g.,][]{1999AJ....117..920A, 1999AcA....49..149P, 2000AcA....50..337P, 2004ApJ...601..260N, 2004A&A...423...97B, 2006A&A...460..459B, 2009AJ....138.1243H, 2010A&A...517A..50G, 2011A&A...527A.116C, 2013A&A...552A.144S, 2014NewA...28...27J, 2016AcA....66..149J, 2016MNRAS.463.1446N, 2016RAA....16...61J, 2018MNRAS.478.5017R, 2018ApJ...866...90C}, among others. For example,  \cite{2004A&A...423...97B, 2006A&A...460..459B} suggested the existence of homogeneous, old and metal-poor stellar halo in the LMC based on the velocity dispersion of RR Lyrae stars. On the basis of spatial and age distributions of Cepheids reported in the Optical Gravitational Lensing Experiment (OGLE) Phase-III survey,  \cite{2014NewA...28...27J} and \cite{2016RAA....16...61J} affirmed that a major star formation triggered in these two dwarf galaxies at $\approx 200\pm50$ Myr ago. \cite{2018ApJ...866...90C} revealed a significant warp in the south west region of LMC outer disk towards SMC and an off-centered tilted LMC bar which they believe is consistent with a direct collision between the two clouds in agreement with the earlier studies \citep[e.g.,][]{2012MNRAS.421.2109B, 2017MNRAS.471.4571C, 2018ApJ...864...55Z, 2019ApJ...874...78Z}. On the basis of VMC data in the near-infrared filters, \cite{2018MNRAS.478.5017R} provided maps for the mass distribution of stars of different ages, and a total mass of stars ever formed in the SMC. These kind of studies are possible due to availability of enormous amount of data generated by the large sky surveys in these directions like Massive Compact Halo Objects Survey \citep{1995AJ....109.1653A}, Optical Gravitational Lensing Experiment \citep{1997AcA....47..319U}, Magellanic Clouds Photometric Survey \citep{1997AJ....114.1002Z}, Two Micron All Sky Survey \citep{2006AJ....131.1163S}, VISTA survey of Magellanic Clouds system \citep{2011A&A...527A.116C}, Survey of the MAgellanic Stellar History \citep{2017AJ....154..199N}, etc.

It has long been observed that reddening (a measurement of the selective total dust extinction) information is one of the important parameter to estimate the structure of the disk of the MCs as well as deriving SFH in these two dwarf galaxies. For example, \cite{2013A&A...552A.144S} suggested that the extra-planer features which are found both in front of and behind the disk could be in the plane of the disk itself if there would be an under-estimate or over-estimate of the extinction values in the direction of the LMC. \cite{2018MNRAS.478.5017R} derived the extinction $A_V$ varying between $\sim$0.1 mag to 0.9 mag across the SMC and found that the high-extinction values follow the distribution of the youngest stellar populations. By simultaneously solving SFH, mean distance, reddening, etc, they were able to get more reliable picture of how the mean distances, extinction values, SF rate, and metallicities vary across the SMC. \cite{2018ApJ...866...90C} presented $E(g-i)$ reddening map across the LMC disk through the colour of red clump (RC) stars and suggested that the majority of the reddening toward the non-central regions results from the Milky Way foreground, which acts as a dust screen on stars behind it. They were able to explore the detailed three-dimensional structure of the LMC and detect a new stellar warp by using their accurate and precise two-dimensional reddening map in the LMC disk.

In our analysis, we use Classical Cepheids or Population I Cepheids (here onward we simply use the term Cepheids for these objects) to map the reddening and probe the recent SFH within the MCs. Cepheids are relatively young massive stars in the core He-burning phase and occupy the space in a well defined instability strip in the H-R diagram. These are regarded as an excellent tracer for the understanding of recent star formation in the host galaxy and one of the most important establishment of the cosmic distance ladder. They play a pivotal role in structure studies of nearby external galaxies like MCs, M31, and M33, among others as they are found in plenty in these galaxies. These pulsating variables, because of their large intrinsic brightness and periodic luminosity variation over time, can be easily detected in the outskirts of the Galactic disk and in nearby galaxies.

Cepheids pulsate in two modes, one called FU mode in which all parts of the system move sinusoidally with the same frequency and a fixed phase relation, and other are called FO Cepheids which pulsate with short period and small amplitude. The distinction between FU and FO Cepheids can easily be made using their positions in the $P$-$L$ diagram. Cepheids pulsating in the FO mode are about 1 mag brighter and relatively lower amplitude than the FU mode for the same pulsation period \citep[e.g.,][]{1999AcA....49..201U, 2005ApJ...621..966B}. One can also distinguish the two modes through the Fourier parameters $R_{21}$ and $\phi_{21}$ determined from the shape of their light curves. During the past few decades, many studies have been directed towards the understanding structure and reddening distribution in the MCs employing the Cepheids \citep[e.g.,][]{1999AcA....49..201U, 2004ApJ...601..260N, 2015AA...573A.135S, 2014NewA...28...27J, 2016RAA....16...61J, 2016ApJ...832..176I, 2016ApJ...816...49S}. 

In the present study, we primarily aim to understand the reddening variation across the MCs through multi-band $P$-$L$ relations of the Cepheids by taking advantage of largest and most homogeneous sample of FU and FO Cepheids spread all across the MCs in the recently released catalogue of OGLE-IV survey. Here, we construct large number of narrow sub-regions in these two galaxies and select significant number of Cepheids in each sub-region to draw respective P-L diagrams. Furthermore, we convert period of FO Cepheids to the corresponding period of FU Cepheids in order to make a single $P$-$L$ diagram with increased sample size of Cepheids. Then using the multi-band $P$-$L$ diagrams in each sub-region, we estimate the corresponding reddening values. In this way, we obtain the reddening in all such sub-regions that are used to construct the reddening map which is of vital importance to study the dust distribution within different regions of these two galaxies. We also estimate the age of Cepheids in order to probe the recent SFH in the two clouds.

This paper is structured as follows. In Section~\ref{data}, we describe the detail of the data used in the present study. The spatial distribution of Cepheids in the MCs is described in Section~\ref{spa_distri}. The reddening determinations and construction of reddening maps are presented in Section~\ref{reddening}. Our analysis on the Cepheids age determination and recent SFH in the MCs is given in Section~\ref{distribution}. The discussion and conclusion of this study are summarized in Section~\ref{conclu}.
\section{Data}\label{data}
There has been plenty of surveys for Cepheids in the MCs, however, OGLE\footnote{http://ogle.astrouw.edu.pl/} has revolutionized the field by producing thousands of Cepheid light curves in the Galaxy, the LMC and the SMC with high signal-to-noise ratio and determining very accurate parameters like periods, magnitudes and amplitudes. OGLE has begun its survey in 1992 in its first phase, followed by three more phases appending additional sky coverage. The fourth phase of OGLE survey has been carried out using a 32-chip mosaic CCD camera on a 1.3-m Warsaw University Telescope at Las Campanas Observatory, Chile between 2010 March and 2015 July which has increased observing capabilities by almost an order of magnitude compared to OGLE-III phase and covered over 3000 square degrees in the sky \citep{2016AcA....66..131S}. The complete detail of the reduction procedures, photometric calibrations and astrometric transformations is available in \cite{2015AcA....65....1U}.

Recently OGLE-IV survey has released high quality photometric data in the $V$ and $I$ band from their observations of Magellanic System \citep{2015AcA....65..233S, 2017AcA....67..103S}. This work utilizes the publicly available photometric catalog of variable stars consisting of 4704 Cepheids in the LMC and 4945 Cepheids in the SMC. The LMC Cepheids sample consists of 2476 FU, 1775 FO, 26 second-overtone (SO), 95 double-mode FU/FO, 322 double-mode FO/SO, 1 double mode FO/Third overtone (TO),1 double mode SO/TO, and 8 triple-mode Cepheids \citep{2015AcA....65..233S}. Similarly, the SMC sample consists of 2753 $FU$, 1793 $FO$, 91 $SO$, 68 $FU/FO$, 239 $FO/SO$, and one triple-mode Cepheids \citep{2017AcA....67..103S}. The sample is reported to be over 99\% complete making it the most complete and least contaminated sample of Cepheids in the Magellanic System \citep{2016AcA....66..149J}. For the present study, we have only used the archival $V$ and $I$ band data containing the mean magnitude and period of FU and FO Cepheids for both the dwarf galaxies. In total, we used simultaneous $V$ and $I$ band data of 4251 Cepheids in the LMC and 4546 Cepheids in the SMC. The photometric uncertainty on OGLE-IV mean magnitudes is found to be $\sigma_{I,V} = 0.007$ mag for brighter and longer period Cepheids ($0.7 \le log~P \le 1.5$) and $\sigma_{I,V} = 0.02$ mag for fainter and shorter period Cepheids ($log~P < 0.7$) \citep{2016ApJ...832..176I}.
\section{Spatial Distribution}\label{spa_distri}
During the last several years, a large number of studies based on different stellar populations have been carried out to understand the structure of the MCs \citep[e.g.,][]{2010A&A...517A..50G, 2013MNRAS.431.1565W, 2013A&A...552A.144S, 2014NewA...28...27J, 2016RAA....16...61J, 2018MNRAS.473.3131M}. In the present study, we have a total of 4251 Cepheids in the LMC and 4546 Cepheids in the SMC, largest sample among any previous studies so far which provides a chance to examine their spatial distributions within the clouds in order to understand the geometry of the disk of these galaxies. To carry out this analysis, we first converted the ($RA$, $DEC$) coordinates to ($X$, $Y$) coordinates using the relations given by \cite{1967pras.book.....V}. Then we divided the LMC in $45\times45$ segments with a dimension of $0.4\times0.4$ kpc square for each segment having a mean spatial resolution of $\approx$ 1.2 deg$^2$. For the SMC, we divided in $90\times90$ segments with a smaller dimension of $0.2\times0.2$ kpc square having a mean spatial resolution of $\approx$ 0.22 deg$^2$ because SMC contains higher density of Cepheids in its relatively small area. We kept the same spatial cell size for all subsequent analysis in our study. The cell size was chosen in such a way that we can get enough Cepheids in each cell for a statistically meaningful determination of mean reddening and recent SFH measurements. Our selection has resulted 389 segments covering a total area of about 456.7 deg$^2$ in the LMC and 562 segments covering a total area of about 131.3 deg$^2$ in the SMC. We counted the total number of Cepheids in each segment in the LMC and SMC, and constructed the spatial map of the Cepheids distribution.
%
%--------------- Figure 1 --------------------------
\begin{figure}[h]
\centering
\includegraphics[width=10.0cm,height=8.0cm]{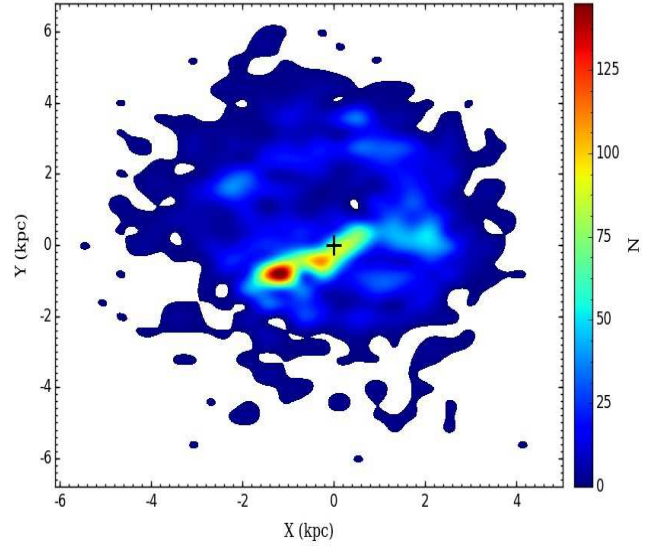}
\caption{Two-dimensional spatial map in the LMC as a function of number of Cepheids measured in different segments. North is up and east is to the left. The location of the optical center of the LMC is shown by plus sign.}
\label{spatial_lmc}
\end{figure}
%--------------- End Figure -----------------------------
 
%--------------- Figure 2 --------------------------
\begin{figure}[h]
\centering
\includegraphics[width=10.0cm,height=8.0cm]{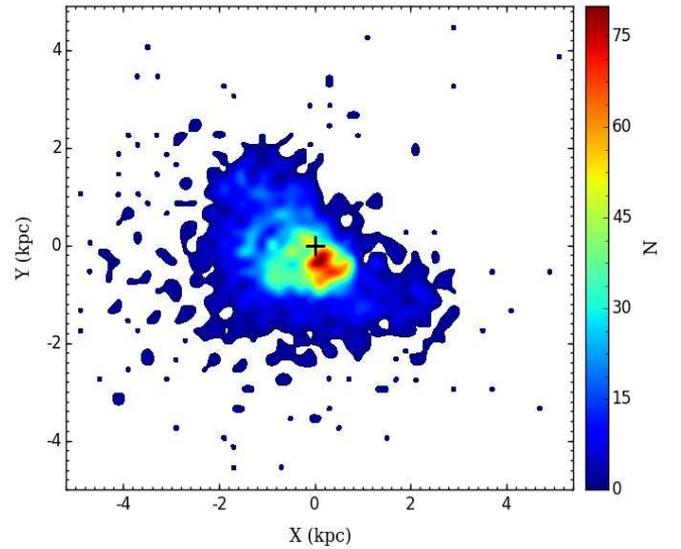}
\caption{Same as Figure~\ref{spatial_lmc} but for the SMC.}
\label{spatial_smc}
\end{figure}
%--------------- End Figure -----------------------------
 
{ In Figure~\ref{spatial_lmc}, we show the two-dimensional spatial distribution of  Cepheids in each of 389 segments in the LMC.} The colour code in the map represents the number of Cepheids in each segment. The observed spatial maps in the LMC shows some interesting structures. The center of the Cepheids density region in the LMC is found to be at $X=-1.2$, $Y=-0.8$ which corresponds to the coordinates $\alpha \sim 84^{o}.00$ and $\delta \sim -70^{o}.29$. It is apparent that the stellar overdensities do not match with the optical center of the LMC but shifted far away towards south-east direction. The spatial distribution of Cepheids along the disk is very distinctive and a bar structure is conspicuous where most of the Cepheids are located. It is also seen that the density structure of the bar region is not smooth. Many Cepheids are distributed in the clumpy structure along the bar which is elongated in the east-west direction where eastern side of the bar shows higher density of Cepheids as compared to its western side. On the basis of RC stars in the LMC, the presence of warp in the bar was also noticed by \cite{2003ApJ...598L..19S} indicating a dynamically disturbed structure for the LMC. The spatial density of Cepheids in the LMC is found to be very poor in its northern arm. 

In Figure~\ref{spatial_smc}, we present similar two-dimensional spatial map for the SMC constructed from 562 segments of our chosen size. As one can see in the map, disk of the SMC seems to be quite irregular and asymmetric. It is evident that the south-western region of the SMC is the most populated and centroid of the Cepheids density distribution is located at $\alpha \sim 12^{o}.73$, $\delta \sim -73^{o}.07$. While dense region is much farther away in the LMC from its optical center, the centroid of the most populated region having Cepheids lies very close to the optical center in the SMC.
\section{Reddening}\label{reddening}
\subsection{Methodology of reddening determination}\label{method}
The dust and gas is heterogeneously distributed within the MCs and a differential extinction is found to exist in these two galaxies hence a constant extinction cannot be applied in deriving $P$-$L$ relations across the LMC and SMC. However, one can study the extinction variation in these galaxies by determining reddening in small sub-regions. Since thousands of Cepheids are available in the MCs, we can divide both the galaxies into small segments containing tens of Cepheids in each zone. For each segment, $P$-$L$ diagram is plotted in the form of
\begin{equation}
m_\lambda = a_\lambda~logP + ZP_\lambda
\label{eqno1} 
\end{equation}
where $m_\lambda$, $a_\lambda$, $P$, and $ZP_\lambda$ denote apparent magnitude, $P$-$L$ slope, period of Cepheid, and zero point, respectively for a given bandpass. Once slope $a_\lambda$ is known through calibrated $P$-$L$ relations, one can obtain the $ZP_\lambda$. 

As OGLE data is given in $V$ and $I$ band and there are many calibrated $P$-$L$ relations for Cepheids in these bands derived in the past studies for the MCs \citep[e.g.,][]{1999AcA....49..201U, 1994MNRAS.266..441L, 2001ApJ...553...47F, 2004A&A...424...43S, 2009A&A...493..471S, 2009ApJ...693..691N, 2015ApJ...808...67N}, we used the most recent ones given by \cite{2009ApJ...693..691N} for the LMC FU Cepheids as following:
\begin{equation}
\ M_{V} = -2.769(\pm 0.023)~logP + 17.115(\pm 0.015)
\label{eqno2}
\end{equation}
\begin{equation}
\ M_{I} = -2.961(\pm 0.015)~logP + 16.629(\pm 0.010)
\label{eqno3}
\end{equation}
For the SMC FU Cepheids, we used the $P$-$L$ relation given by \cite{2015ApJ...808...67N} as following:
\begin{equation}
\ M_{V} = -2.660(\pm 0.040)~logP + 17.606(\pm 0.028)
\label{eqno4}
\end{equation}
\begin{equation}
\ M_{I} = -2.918(\pm 0.031)~logP + 17.127(\pm 0.022)
\label{eqno5}
\end{equation}
\noindent where $M_{V}$ and $M_{I}$ denote absolute magnitudes in $V$ and $I$ band, respectively. We note here that no significant metallicity gradient has been seen across the LMC and SMC  \citep[e.g.,][]{2006AJ....132.1630G, 2009A&A...506.1137C, 2009AJ....138..517P, 2010MNRAS.408L..76F, 2014MNRAS.438.2440D, 2016RAA....16...61J} hence any variation in the MCs $P$-$L$ relations due to metallicity variation is not considered in the present study. Even if we accept any variation in the metallicity at all as reported in some previous studies \citep[e.g.,][]{2003A&A...404..423T, 2008AJ....136.1039C, 2012MNRAS.426.2063K, 2014MNRAS.442.1680D} the difference in reddening is found to be $\sim$0.001 mag which is insignificant in comparison of the scatter in the $P$-$L$ diagrams itself. Moreover, there are many studies reported the non-linearity in the $P$-$L$ relations in these galaxies \citep{1999A&A...348..175E, 1999AcA....49..201U, 2002AJ....123.3216S, 2004MNRAS.350..962K, 2009A&A...493..471S, 2010AcA....60...17S, 2011A&A...531A.134T, 2015AA...573A.135S, 2016MNRAS.458.3705B}. A wide range of break points has been reported at different places in the $P$-$L$ diagrams (1.0, 2.5, 2.9, 3.55, and 10~day) which also varies in different bands and different overtones. However, we did not consider any breaks in the $P$-$L$ diagrams because uncertainties from photometry and low number statistics in the present data were larger than the uncertainties from breaks. With these assumptions, we determine the reddening in large number of segments across these two galaxies.

%
%--------------- Figure 3 -----------------------------------
\begin{figure}[h]
\centering
\hspace*{-0.3cm}\includegraphics[width=10.3cm,height=10.0cm]{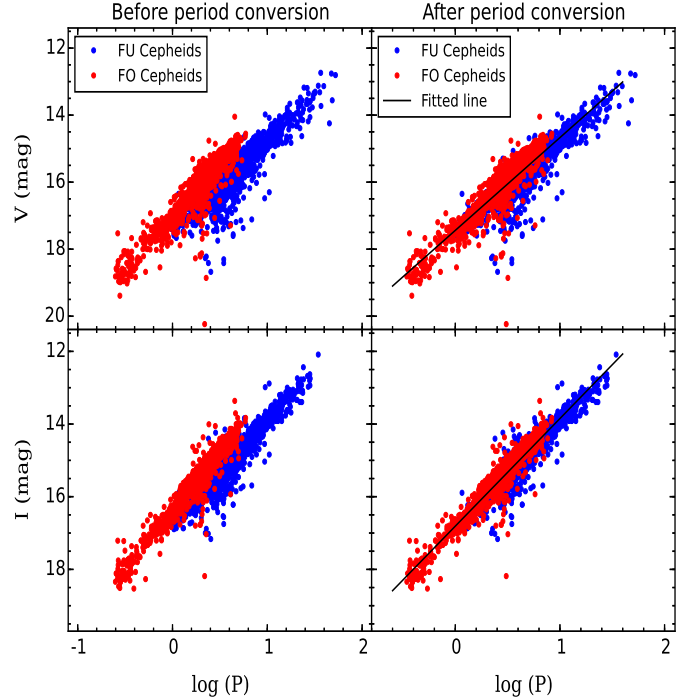}
\caption{The $P$-$L$ diagrams in the LMC for the FU and FO Cepheids in the left panels. The right panels show the combined FU and FO Cepheids after converting period of FO Cepheids to corresponding periods of FU Cepheids. Upper panels are for $V$ filter and lower panels are for $I$ filter. Points shown in blue color are for FU Cepheids and red color are for FO Cepheids.}
\label{pl_lmc}
\end{figure}
%--------------- End Figure ---------------------

It has been however observed through the $P$-$L$ diagram of FU and FO Cepheids that for the same pulsation period FU Cepheids have greater magnitude as compared to FO Cepheids and they follow two different $P$-$L$ relations. Hence, to combine the FU and FO Cepheids together in a single $P$-$L$ diagram, one requires to first convert period of FO Cepheids into the corresponding period of FU Cepheids which requires a well defined relation connecting the two periods. As vast majority of Cepheids pulsates only in a single mode, there are only a limited number of Cepheids which pulsate in two modes simultaneously \citep{2018A&A...618A.160L}. Using the high resolution spectroscopic observations of 17 such double mode Galactic Cepheids, \cite{2007A&A...473..579S} derived a relation for transformation of FO and FU periods as following:
\begin{equation}
\frac{P_{1}}{P_{0}} = -0.0143~logP_{0}-0.0265~\left[\frac{Fe}{H}\right]+0.7101
\label{eqno6} 
\end{equation}
$~~~~~~~~~\pm0.0025~~~~~~~~~\pm0.0044~~~~~~~~~~~\pm0.0014$\\

\noindent where $P_{1}$ and $P_{0}$ denote period of FO Cepheid and corresponding period of FU Cepheid, respectively. $\left[\frac{Fe}{H}\right]$ denotes the metallicity. We converted periods of FO Cepheids to those of the corresponding periods of FU Cepheids to draw $P$-$L$ diagrams of combined sample of FU and FO Cepheids. Here, we considered a mean present-day metallicity of $-0.34\pm0.03$ dex for the LMC based on Cepheids \citep{2006ApJ...642..834K} and $-0.70\pm0.07$ dex for the SMC based on supergiants \citep{1997A&A...323..461H, 1999ApJ...518..405V}.

%--------------- Figure 4 -----------------------------------
\begin{figure}[h]
\centering
\hspace*{-0.3cm}\includegraphics[width=10.3cm,height=10.0cm]{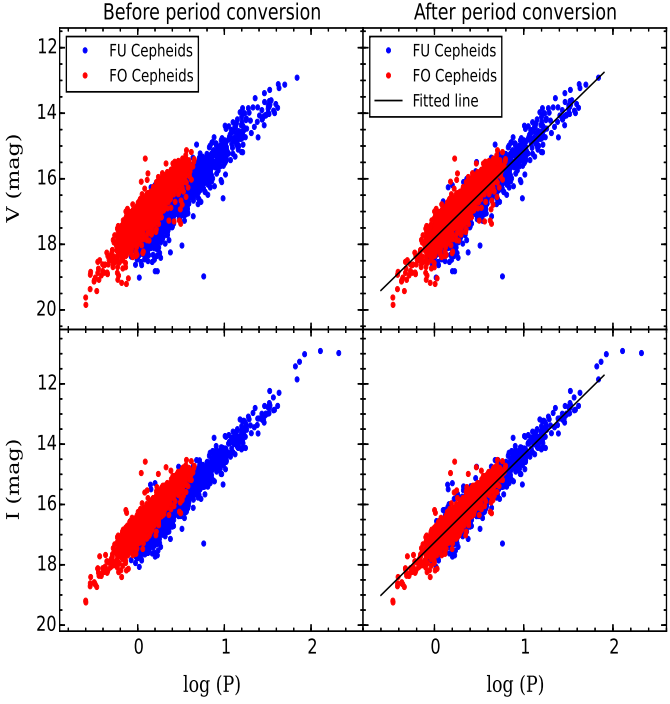}
\caption{Same as Figure~\ref{pl_lmc} but for the SMC.}
\label{pl_smc}
\end{figure}
%--------------- End Figure --------------------
%

In Figures~\ref{pl_lmc} and \ref{pl_smc}, we illustrate $V$ and $I$ band $P$-$L$ diagrams in the LMC and SMC, respectively. In each diagram, we show FU and FO Cepheids independently in the left panel and combined FU and FO Cepheids in the right panel after the period conversion. It is conspicuous that FU and FO Cepheids follow two different $P$-$L$ relations and after converting FO periods to corresponding FU periods using eq.~\ref{eqno6}, they fall on a single $P$-$L$ relation in their respective filters in both the LMC and SMC, as can be seen in the right plots of Figures~\ref{pl_lmc} and \ref{pl_smc}. This approach allows us to combine two different modes of Cepheids in a single P-L relation which in turn gives us a larger sample of Cepheids for the study. We then draw multi-band period versus magnitude diagrams in each segment and fit the calibrated $P$-$L$ relation to determine the corresponding intercept. The difference in intercepts between calibrated and observed $P$-$L$ relations yields the $ZP$ value for a given passband. 

\begin{equation}
ZP = ZP_{\sl observed} - ZP_{\sl calibrated}
\label{eqno7} 
\end{equation}

\noindent Here, $ZP_{calibrated}$ in the $V$ and $I$ bands are respectively taken as 17.115 mag and 16.629 mag for the LMC and 17.606 and 17.127 for the SMC as given in the eqs.~2 to 5. The $ZP_{observed}$ varies for each segment mainly according to their relative dust extinction, whereas $P$-$L$ slope is almost fixed in a galaxy. Once we know the $ZP$ values in two different passbands, we derive the reddening in the selected region by subtracting them as follows:
\begin{equation}
E(V-I) = ZP_V - ZP_I
\label{eqno8} 
\end{equation}
where $ZP_V$ and $ZP_I$ are the zero points in $V$ and $I$ band, respectively. 

Assuming that the MCs Cepheids follow the same reddening law as that of Galactic Cepheids, we can determine reddening $E(B-V)$ using the following reddening ratio \citep{2017MNRAS.466.4138W}:
\begin{equation}
 E(B-V) = E(V-I)/1.32
\label{eqno9} 
\end{equation}
In this way, one can estimate reddening $E(V-I)$ and $E(B-V)$ in each segment of the LMC and SMC. The uncertainties in $E(V-I)$ and $E(B-V))$ values are calculated by combining the error in the $ZP$ values and those in the P-L slopes. It is important to note here that one does not require any prior information on the distance of the host galaxy of Cepheids to map the reddening when their multi-band photometry is available. Since we have $V$ and $I$ band photometry for thousands of Cepheids in the MCs, it ideally suited us to study the reddening in detail in these two nearby clouds which is crucial to understand the dust distribution and probe the recent SFH in the MCs.
\subsection{Reddening in the MCs}\label{extinct}
Although one can measure reddening corresponds to each Cepheid but the mean magnitude of individual Cepheid determined through the photometric light curves may contain some uncertainty which propagates in the reddening estimation. Therefore, to carry out analysis, we made small segments in the LMC and SMC as defined in Section~\ref{spa_distri}. To draw $P$-$L$ relations, we need to select those segments which have a larger number of Cepheids lying in the same direction in order to determine more precise value of mean reddening with lesser uncertainty in any given direction. Therefore only those segments were considered which contain a minimum of 10 Cepheids (3 $\times$ Poissonian error) in at least one passband.

\subsubsection{The LMC}\label{Redlmc}
We selected a total 133 segments with an average angular resolution of $\approx$ 1.2 deg$^2$ covering a total area of about 154.6 deg$^2$ in the LMC. The maximum number of Cepheids among selected segments is found to be 144 in the LMC. Keeping the slopes fixed in the $V$ and $I$ band $P$-$L$ diagrams as given in eqs.~\ref{eqno2} and \ref{eqno3}, we estimated $ZP$ values for each segment using a least-square fit to the observed period versus magnitude diagrams. It should be noted here that uncertainties in the measurements of apparent $V$ and $I$ magnitudes are not provided in the OGLE-IV catalogue hence individual photometric errors are not considered in the fit.

%--------------- Figure 5 -----------------------------------
\begin{figure}[h]
\centering
\hspace*{-0.3cm}\includegraphics[width=10.0cm,height=10.0cm]{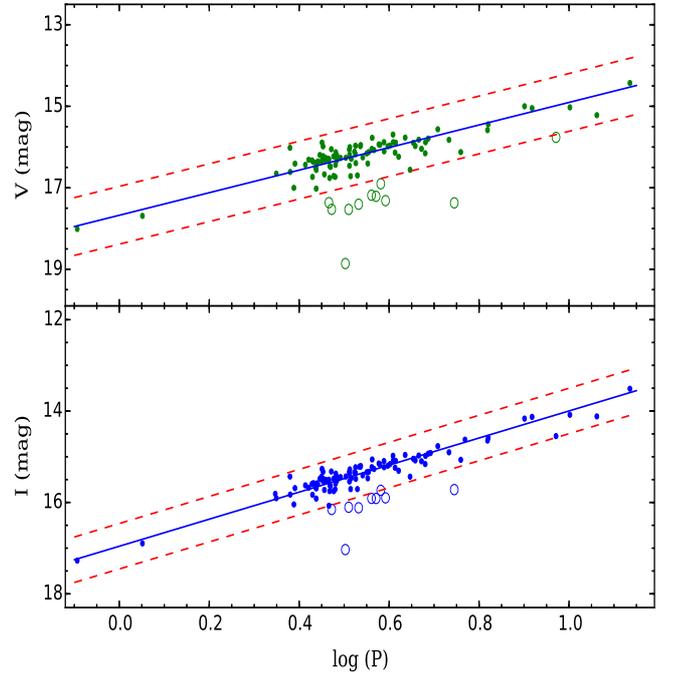}
\caption{The $P$-$L$ relations for a sub-region in the LMC in $V$ band (upper panel) and $I$ band (lower panel). Here, the continuous lines represent best linear fit with fixed slope of -2.769 for $V$ band and -2.961 for $I$ band, respectively. The dashed lines represent the 3$\sigma$ cut lines and Cepheids outside 3$\sigma$ lines are shown by open circles which are not considered in the best fit.}
\label{PLdia_lmc}
\end{figure}
%--------------- End Figure ------------------------ 
%
 
%------------------------------------ Table 1 -------------------------
\begin{table}[h]
  \centering
  \caption{Table provides reddening in 133 different segments of the LMC. Here, Ist column represents ID of the segment, 2nd and 3rd columns give central (X,Y) coordinate, 4th and 5th columns give corresponding (RA,DEC) coordinates. The last 3 columns give reddening $E(V-I)$, $E(B-V)$ and error in the $E(V-I)$.}
\tiny
  \label{tab:ext_lmc}
  \begin{tabular}{cccccccc}
  \hline 
    Seg & $X_c$ & $Y_c$  & $RA_c$ & $DEC_c$ & $E(V-I)$ & $E(B-V)$ & Error\\
         ID & (kpc)  &  (kpc)  & (deg) & (deg) & (mag)    &  (mag) &  (mag)\\
\hline \\
001 & -0.8 & -2.0 & 82.21 & -71.25 & 0.124 & 0.094 & 0.053\\
002 & -0.4 & -2.0 & 81.06 & -71.26 & 0.116 & 0.088 & 0.058\\
003 & 0.4 & -2.0 & 78.76 & -71.26 & 0.088 & 0.067 & 0.071\\
   . & . & . & . & . & . & . & . \\
   . & . & . & . & . & . & . & . \\
   . & . & . & . & . & . & . & . \\
133 & 0.4 & 4.0 & 79.01 & -65.74 & 0.123 & 0.093 & 0.092\\
  \hline
 \end{tabular}
\end{table}
%---------------------------------------------------------------

\begin{figure}[h]
\centering
\hspace*{-0.3cm}\includegraphics[width=10.0cm,height=7.0cm]{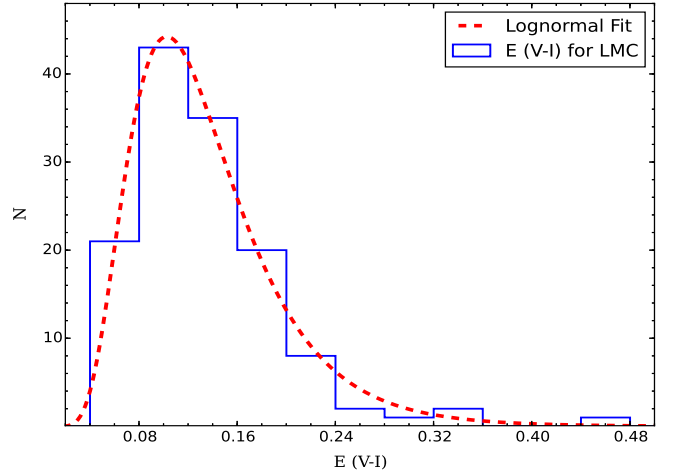}
\caption{Histogram of reddening values in 133 segments of the LMC shown by continuous line. The dashed line represents the best fit with a log-normal distribution.} 
\label{hist_lmc}
\end{figure}

In Figure~\ref{PLdia_lmc}, we show one such randomly selected $P$-$L$ diagram, each in $V$ and $I$ band. We here apply $3\sigma$ cut to reduce the contamination in iterative manner unless all the Cepheids fall within $3\sigma$ lines of the given $P$-$L$ slope where $\sigma$ is the rms derived from fitting $P$-$L$ relation to the observed data points. In Figure~\ref{PLdia_lmc}, the best fit slope is drawn by continuous line while final 3-$\sigma$ deviation is shown by dashed lines on both sides of the best fit. It should be noted here that some intrinsic dispersion in the $P$-$L$ diagram is obvious due to the finite width of the Cepheid instability strip as well as contamination due to blending and crowding effects  \citep{1968ApJ...151..531S, 2003A&A...402..113J, 2010A&A...512A..66J}. On an average the dispersion in the LMC Cepheid $P$-$L$ diagrams is found to be $\lesssim 0.06$ mag in both $V$ and $I$ band. We note here that an uncertainty in the LMC metallicity of 0.03 dex \citep{2006ApJ...642..834K} used in eq.~\ref{eqno6} gives only an added 0.001 mag error in the $ZP$ values determined through combined (FU+FO) $P$-$L$ relations which is much smaller than the typical error on the $ZP$ estimations. Using the zero point intercepts in $V$ and $I$ band, we determined the reddening $E(V-I)$ and $E(B-V)$ values as described in eqs.~\ref{eqno8} and ~\ref{eqno9} for all the 133 segments in the LMC. Table~\ref{tab:ext_lmc} lists the central coordinates of all the selected segments in the LMC, corresponding to which reddening values $E(V-I)$ and $E(B-V)$ are given in columns 6 and 7. The complete table is available online or from the lead author. The value of reddening $E(V-I)$ in different segments of the LMC varies from 0.041 mag to 0.466 mag with a mean value of $0.134\pm0.006$ mag. The histogram shown in Figure~\ref{hist_lmc} illustrates the distribution of reddening $E(V-I)$ in the LMC.
%-------------------------------------Table 2 ------------------------------
\begin{table*}[h]
  \centering
\vspace{-0.2cm}
  \caption{A summary of recent reddening measurements using different stellar populations in the LMC.}
  \label{tab:comp_lmc}
  \begin{tabular}{c|c|c|c}
\hline
E(V-I) & E(B-V) & Tracer used for the estimate & Reference \\
(mag)   &  (mag)   &                              &           \\
\hline
                  & $0.09\pm 0.02$   & B Stars & \cite{2000AA...364..455L} \\
                  & $0.06$           & MC's spectra & \cite{2001AA...371..895D} \\
                  & $0.14\pm 0.02$   & Cepheids & \cite{2004ApJ...601..260N} \\
   $0.08\pm 0.04$ &                  & RC stars & \cite{2005AA...430..421S} \\
   $0.09\pm 0.07$ &                  & RC stars & \cite{2011AJ....141..158H} \\
   $0.11\pm 0.06$ &                  & RR Lyrae & \cite{2011AJ....141..158H} \\
   $0.12\pm 0.05$ &                  & RRab     & \cite{2013MNRAS.431.1565W} \\
   $0.11\pm 0.06$ &                  & Clusters & \cite{2013MNRAS.431.1565W} \\
                  &0.10-0.16$\pm 0.02$& Detached EBs & \cite{2013Natur.495...76P} \\
   $0.11\pm 0.09$ &                  & Cepheids & \cite{2016ApJ...832..176I} \\
                  & $0.093\pm 0.031$ & RC stars & \cite{2018ApJ...866...90C} \\
   $0.113\pm 0.060$& $0.091\pm 0.050$& FU and FO Cepheids   & Present study \\
\hline
\end{tabular}
\end{table*}
%-----------------------------------------------------------------------------------------

The probability distribution function of interstellar column density measurements is known to be close to log-normal in the solar neighborhood \citep[e.g.,][]{2010A&A...512A..67L}. Hence we draw a log-normal profile in the reddening distribution of the LMC. The best profile illustrated by dashed line is shown in Figure~\ref{hist_lmc} that shows it matches the distribution with exquisite accuracy. The uncertainty in the reddening $E(V-I)$ was determined from the $\sigma$-value in the best fit profile and error in the $ZP$ values in $V$ and $I$ band. First, we calculated error in each value of $ZP$ using the uncertainties in the slope and zero points of the $P$-$L$ relation in both the filters. Then, combined error in the estimation of $E(V-I)$ has been determined through the error propagation
$$\sigma(E(V-I)) = \sqrt{(\sigma(ZP_V)^2 + \sigma(ZP_I)^2 + \sigma(fit)^2}$$
\noindent We determined error in each value of $E(V-I)$ that is given in the last column of Table 1. A mean error in the $E(V-I)$ values was taken corresponding to the peak in the distribution of $\sigma(E(V-I))$ values obtained in all the 136 segments. The mean value of reddening using the log-normal profile fit is found to be $E(V-I)=0.113\pm0.060$ mag. The corresponding mean value of $E(B-V)$ is estimated as $0.091\pm0.050$ mag.

%--------------- Figure 6 -------------------------------
%--------------- End Figure --------------------%
%

The reddening in the LMC has been estimated by several authors in the past and we summarize some of the recent reddening measures in $E(B-V)$ or $E(V-I)$ using different stellar populations in Table~\ref{tab:comp_lmc}. Among some of the most recent estimates of reddening in the LMC, the RC stars used by \cite{2011AJ....141..158H} indicated a mean reddening of the LMC as $E(V-I)=0.09\pm0.07$ mag while with RR Lyrae stars a median value of $E(V-I)=0.11\pm0.06$ mag was obtained. \cite{2016ApJ...832..176I} provided reddening estimates derived through Cepheids at 7 different positions across the LMC and found that $E(B-V)$ varies in the range of $0.08\pm0.03$ mag to $0.12\pm0.02$ mag. Furthermore, a broader range of $E(B-V)$ from $0.10\pm0.02$ mag to $0.16\pm0.02$ mag was given by \cite{2013Natur.495...76P} using detached eclipsing binaries found in the OGLE data. \cite{2018ApJ...866...90C} constructed 2D reddening map of LMC disk with help of red clump (RC) stars observed in the Survey of the MAgellanic Stellar History (SMASH) and reported an average $E(g-i)=0.15\pm0.05$ mag. Using the conversion equations from \cite{2018ApJS..239...18A}, this reddening corresponds to $E(B-V)=0.093\pm0.031$ mag. From the Table~\ref{tab:comp_lmc}, it is found that the mean LMC reddening varies from $E(B-V) \approx 0.031$ mag to $E(B-V) \approx 0.353$ mag. The average LMC reddening $E(V-I)=0.113\pm0.060$ mag and $E(B-V)=0.091\pm0.050$ mag estimated in the present study is in good agreement with the recently reported average reddening measurements using stellar populations of different ages in the LMC.
\subsubsection{The SMC}\label{Redsmc}
\begin{figure}[h]
\centering
\hspace*{-0.3cm}\includegraphics[width=10.0cm,height=10.0cm]{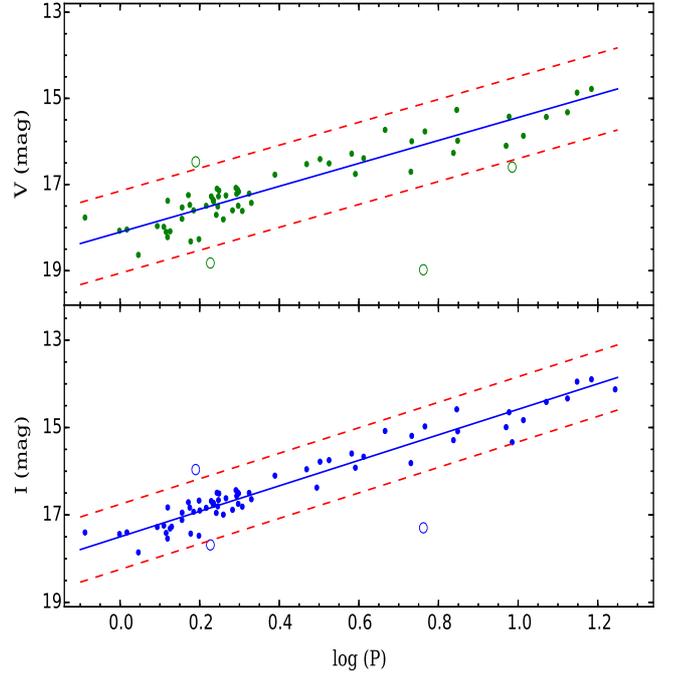}
\caption{Same as Figure~\ref{PLdia_lmc} but for the SMC with fixed slope of -2.66 for $V$ band and -2.918 for $I$ band, respectively .}
\label{PLdia_smc}
\end{figure}

Employing the same approach, we selected 136 segments in the SMC with an average angular resolution of $\approx$ 0.22 deg$^2$ covering a total area of about 31.3 deg$^2$. The maximum number of Cepheids among selected segments is found to be 80 in the SMC. We plot period versus magnitude diagrams for each segment and drawn a least-square linear fit keeping the slopes fixed in the $V$ and $I$ band $P$-$L$ diagrams as given in eqs.~\ref{eqno4} and \ref{eqno5}. We also applied here $3\sigma$ iterative rejection criteria to remove the outliers. In Figure~\ref{PLdia_smc}, we show one such randomly selected $P$-$L$ diagram along with the best fit line, both in $V$ and $I$ band. We also illustrate 3-$\sigma$ cut lines on both sides of the best fit to represent the acceptable deviations in the $P$-$L$ relation. Employing the same approach as that of the LMC, we determined the reddening values of $E(V-I)$ and $E(B-V)$ in each of 136 segments in the SMC. Here, we found 9 segments (6.6\%) have unphysical negative values of reddening and we assigned them zero reddening within the given uncertainties. It should also be noted here that the uncertainty in the SMC metallicity of 0.07 dex \citep{1997A&A...323..461H, 1999ApJ...518..405V} used in eq.~\ref{eqno6} results only an additional 0.001 mag error in the $ZP$ values which is negligible in comparison of the combined uncertainty in the reddening estimates.

\begin{figure}[h]
\centering
\hspace*{-0.3cm}\includegraphics[width=10.0cm,height=7.0cm]{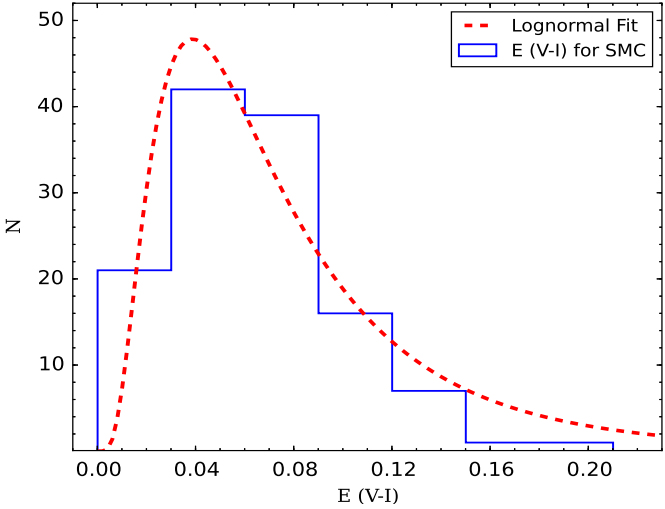}
\caption{Same as Figure~\ref{hist_lmc} but for the 136 segments of the SMC.} 
\label{hist_smc}
\end{figure}

Table~\ref{tab:ext_smc} provides reddening values $E(V-I)$ and corresponding $E(B-V)$ in the columns 6 and 7 for each segment in the SMC. The complete table is available online or from the lead author. The value of reddening $E(V-I)$ in different segments within the SMC varies from 0.0 to 0.189 mag with a mean value of $0.064\pm0.008$ mag. The histogram shown in Figure~\ref{hist_smc} illustrates the distribution of reddening $E(V-I)$ across the SMC. After fitting log-normal profile in the histogram, mean value of reddening is estimated to be $E(V-I) = 0.049\pm0.070$ mag. The equivalent reddening $E(B-V)$ in the SMC is estimated to be $0.038\pm0.053$ mag.

%
%--------------- Figure 7 -----------------------------------
%--------------- End Figure ------------------------
%
%--------------- Figure 8 -------------------------------
%--------------- End Figure --------------------

%----------------------------------------Table 3--------------------------
\begin{table}[h]
  \centering
  \caption{Same as Table~\ref{tab:ext_lmc} but for the 136 degments of the SMC.}
\tiny
  \label{tab:ext_smc}
  \begin{tabular}{cccccccc}
  \hline 
    Seg     & $X_c$ & $Y_c$  & $RA_c$ & $DEC_c$ & $E(V-I)$ & $E(B-V)$ & Error\\
         ID & (kpc)  &  (kpc)  & (deg) & (deg) & (mag)    &  (mag) &  (mag)\\
\hline \\
1 & 0.9 & -1.3 & 10.04 & -73.97 & 0.032 & 0.024 & 0.131\\
2 & 1.1 & -1.3 & 9.37 & -73.96 & 0.055 & 0.042 & 0.089\\
3 & -1.5 & -1.1 & 18.00 & -73.75 & 0.019 & 0.015 & 0.150\\
   . & . & . & . & . & . & . & .\\
   . & . & . & . & . & . & . & .\\
   . & . & . & . & . & . & . & .\\
136 & -1.1 & 1.5 & 16.23 & -71.38 & -0.011 & -0.008 & 0.067\\
  \hline
 \end{tabular}
\end{table}
%------------------------------------------------------------------
%
%--------------------------------------------- Table 4 -------------
\begin{table*}
  \centering
  \caption{A summary of recent reddening measurements using different stellar populations in the SMC.}
  \label{tab:comp_smc}
  \begin{tabular}{c|c|c|c}
\hline
E(V-I) & E(B-V) & Tracer used for the estimate & Reference \\
     (mag)   &  (mag)   &                              &           \\
\hline
                   & $0.037$           & Dust emission  & \cite{1998ApJ...500..525S} \\
                   & $0.07\pm 0.02$    & B Stars & \cite{2000AA...364..455L} \\
                   & $0.04$           & MC's spectra & \cite{2001AA...371..895D} \\
   $0.07\pm 0.06$  &                  & RRab stars & \cite{2011AJ....141..158H} \\
   $0.04\pm 0.06$  &                  & RC stars & \cite{2011AJ....141..158H} \\
   $0.053\pm 0.017$& $0.038$          & RC stars, RR Lyrae & \cite{2012ApJ...744..128S} \\
                   & $0.048\pm 0.039$ & RRab Stars & \cite{2014MNRAS.438.2440D} \\
                   & $0.096\pm 0.080$ & Cepheids & \cite{2015AA...573A.135S} \\
                   & $0.071\pm 0.004$  & Cepheids & \cite{2016ApJ...816...49S} \\
                   & $>$0.08          & YJKs col-mag diagrams    & \cite{2018MNRAS.478.5017R} \\
   $0.06\pm 0.06$  &                  & RR Lyrae & \cite{2018MNRAS.473.3131M} \\        
   $0.049\pm 0.070$ & $0.038\pm 0.053$  & FU and FO Cepheids & Present study\\
\hline
\end{tabular}
\end{table*}
%----------------------------------------------------------------
%
The reddening in the SMC has been studied by several authors in the past through reddening measures in $E(B-V)$ or $E(V-I)$ using different tracers and we provide mean reddening values obtained in some of the recent studies in Table~\ref{tab:comp_smc}. Among the most recent analysis, \cite{2011AJ....141..158H} obtained the mean value of $E(V-I)=0.07\pm0.06$ mag through 1529 RRab stars of OGLE-III survey and $E(V-I)=0.04\pm0.06$ mag using RC stars. A mean reddening of $E(B-V)=0.071\pm0.004$ mag was reported by \cite{2016ApJ...816...49S} using the sample of 92 SMC Cepheids with period greater than 6 days. \cite{2018MNRAS.473.3131M} estimated a reddening of $E(V-I) = 0.06\pm0.06$ mag using the VISTA and OGLE-IV survey data of 2997 fundamental mode RR Lyrae stars. Using infrared colour-magnitude diagrams on the deep VISTA survey data, \cite{2018MNRAS.478.5017R} inferred a large extinction in the central region of the SMC, however, found a relatively lower extinction in the external regions of the SMC. If we compare our present estimates of $E(V-I)=0.049\pm0.070$ mag and $E(B-V)=0.038\pm0.053$ mag  with the recent reddening values given in Table~\ref{tab:comp_smc}, we find that our analysis is in broad agreement with the previous studies except that of \cite{2015AA...573A.135S} which yield higher reddening value of $E(B-V)=0.096\pm0.080$ mag from their study of SMC Cepheids. It is also observed from the Tables \ref{tab:comp_lmc} and \ref{tab:comp_smc} that the mean reddening values obtained in the MCs through different stellar populations do not show any significant variation among different studies.
\subsection{Reddening Maps in the LMC and SMC}\label{Redmap}
The reddening maps are often generated to remove the effects of reddening from the optical and UV images of different regions of the galaxies. They also have wide implications in understanding the star formation rates and recent SFH in the galaxies. In the following subsections, we individually focus our attention to the reddening distributions across the LMC and SMC.
\subsubsection{The LMC}\label{lmc}
\begin{figure}[h]
\centering
\includegraphics[width=10.0cm,height=7.0cm]{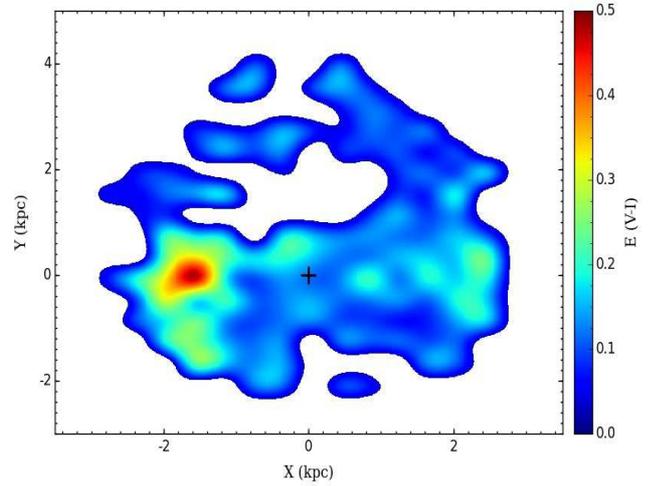}
\caption{The reddening maps emanated through the $P$-$L$ diagrams of the Cepheids in the LMC. North is up and east is to the left. The optical center of the LMC is shown by plus sign. The colour bar represents the interpolated reddening $E(V-I)$.}
\label{map_lmc}
\end{figure}

An examination of individual estimates of reddening values determined through $P$-$L$ relations, we found that the reddening varies from one region to another region of the cloud. While in some regions of the LMC, reddening is negligibly small but in some regions it is found to be significantly high. The reddening values estimated for the 133 selected segments were used to construct a reddening map on a rectangular grid using the central $(X,Y)$ of all segments. Figure~\ref{map_lmc} presents the reddening map derived through the $P$-$L$ diagrams of the  Cepheids, showing the mean reddening along the line of sight toward each segment. Here, a smoothed reddening distribution was adopted to construct the map. The optical center of LMC ($\alpha = 05^h 19^m 38^s \equiv 79^{o}.91$ and $\delta = -69^{o} 27^{'} 5^{''}.2 \equiv -69^{o}.45$; \cite{1972VA.....14..163D}) is shown in the figure. The quality of reddening map is basically regulated by the size of the segments as well as number of Cepheids available in these regions. The sparse spatial distributions of the LMC Cepheids in the outer regions, as shown in Figure~\ref{spatial_lmc}, lend a poor resolution to the reddening map in the outer regions of the LMC as seen in Figure~\ref{map_lmc}.
%
%--------------- Figure 9 -------------------------
%--------------- End Figure -------------------------------

The reddening map of the LMC exhibits a non-uniform distribution of dust across the LMC. With respect to the optical center in the LMC, a lopsidedness in the reddening distribution towards the eastern regions is noticed. While the central region of the LMC contains low values of reddening having $E(V-I)=0.113\pm0.045$ mag, the most promising region in the LMC reddening map is seen towards the north-east region of the LMC outer disk centred at $X=-1.6$ kpc, $Y=0.0$ kpc from the center, which corresponds to $\alpha \sim 85^{o}.13,~\delta \sim -69^{o}.34$ having $E(V-I) = 0.466$ mag. Interestingly, the adjacent region centred around $\alpha \sim 84^o$, $\delta \sim -70^o$ contains the highest concentration of Cepheids in our sample. This region is most likely associated with the star forming HII region 30 Doradus (Tarantula Nebula or NGC 2070) centred at $\alpha = 84^o.5, \delta = -69^o.1$ \citep{2000A&A...355L..27H}. This is the most active star forming region in the bar of the LMC \citep{1999AJ....118.2797K, 2013A&A...554A..33T}. The same region was identified as having highest reddening by \cite{2004ApJ...601..260N} and \cite{2016ApJ...832..176I} as well as having highest concentration of young cluster populations by \cite{2010A&A...517A..50G}. This region is lying in the direction of compact massive cluster RMC 136a that contains a large concentration of young star clusters \citep{2010A&A...517A..50G, 2011A&A...530A.108E}. \cite{2017NatAs...1..784O} found that the 30 Dor is located in the region where the LMC bar joins the H~I arms and such locations are prone to enhance star formation activity due to high concentrations of gas and to the shocks induced by the internal dynamical processes \citep{2018ApJ...853..104B}. In the bar region across the LMC disk, reddening is relatively low and a mean value of $E(V-I) = 0.153$ mag was estimated. On the other side of the LMC bar, few extincted regions are found having moderate reddening. The reddening map, in general, is in good agreement with the maps of HI column density \citep{1992A&A...263...41L} and MACHO Cepheids \citep{2004ApJ...601..260N}. The geometry of the LMC disk was studied by \cite{2004ApJ...601..260N} using the MACHO and 2MASS Cepheids data and estimated the mean reddening $E(B-V)$ by three different methods. All the three estimated reddening values were consistent with each other and the variance-weighted average reddening was found to be $E(B-V)=0.14\pm0.02$ mag. 

\begin{figure}
\centering
\includegraphics[width=10.0cm,height=6.0cm]{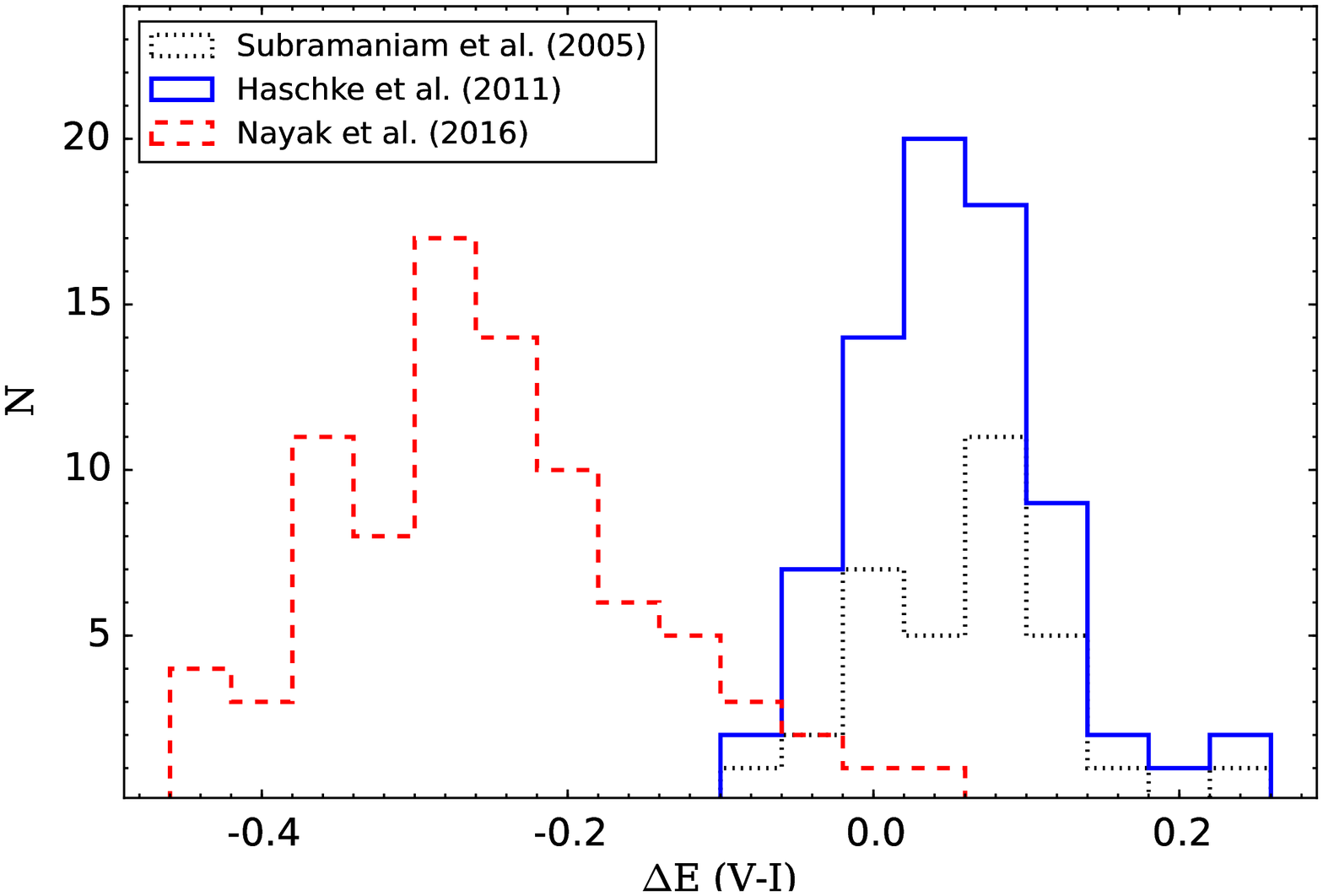}
\includegraphics[width=10.0cm,height=6.0cm]{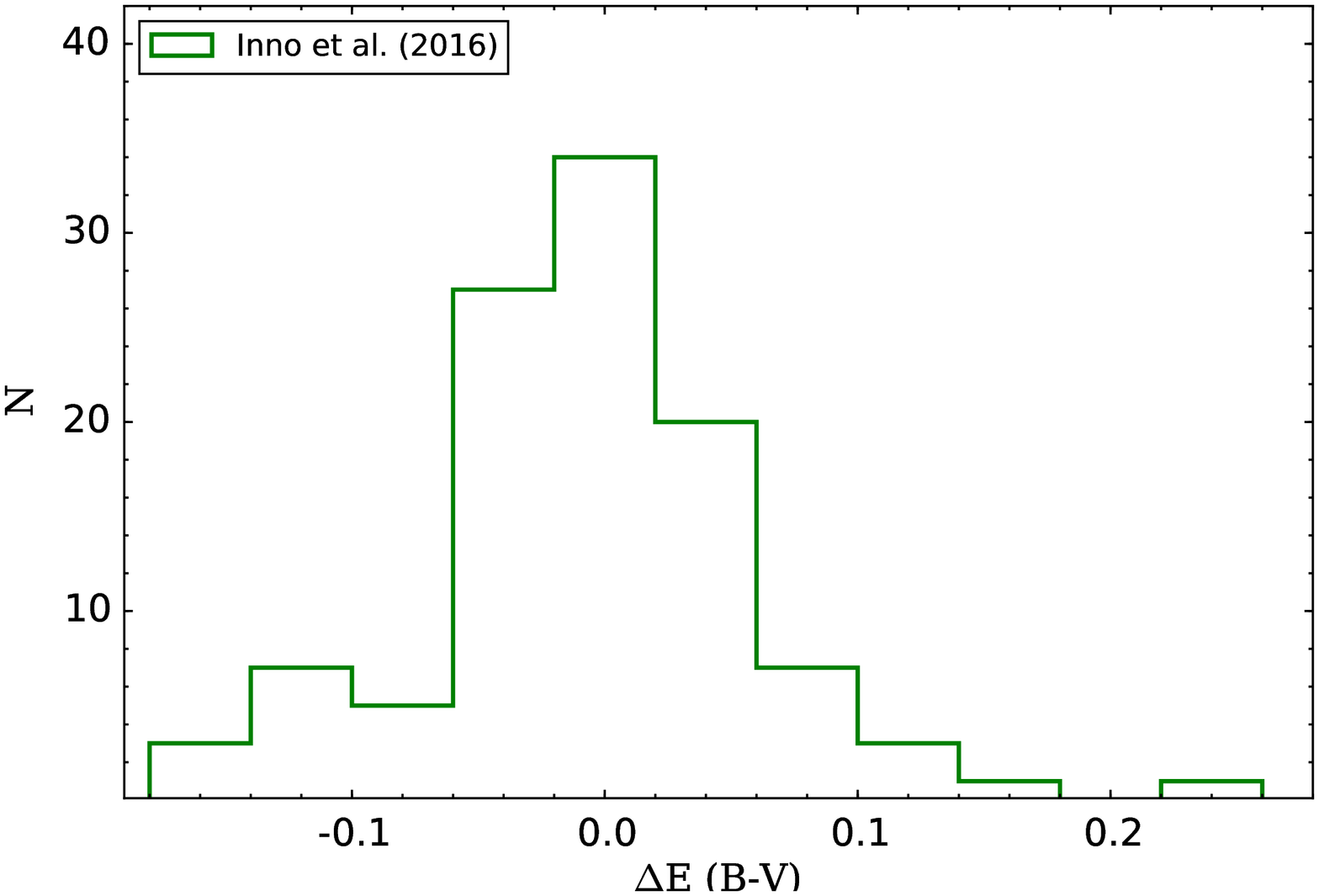}
\caption{(a) Upper panel shows the distributions of the $\Delta E(V-I)$ i.e. past studies results subtracted from our results in the LMC as mentioned on the top-left corner, (b) Lower panel shows the difference $\Delta E(B-V)$ between our reddening estimates and those of \cite{2016ApJ...832..176I}.}
\label{comp_lmc}
\end{figure}

It is found that the reddening maps of the LMC have been constructed in numerous studies in the past using different kind of stellar populations. In order to compare our reddening map with some of the recent studies in the LMC, we took advantage of four such reddening maps in the LMC constructed in the optical bands for which archival data is available \citep[cf.,][]{2005AA...430..421S, 2011AJ....141..158H, 2016MNRAS.463.1446N, 2016ApJ...832..176I}. We performed a cell-by-cell comparison of our reddening map with these maps after cross-examinations of central ($\alpha$, $\delta$) values of each segment with the reddening values given in same region in these maps. The histograms of the comparisons between our reddening map and these maps are illustrated in Figure~\ref{comp_lmc}. We discuss these comparisons in some detail as follows.
\begin{enumerate}
\item The reddening $E(V-I)$ values in the LMC is estimated by \cite{2005AA...430..421S} for 1123 locations and found $E(V-I)$ in the LMC varies between 0.1 to 0.3 mag. Her study shows the total average LMC bar reddening is $E(V-I)=0.08\pm0.04$ mag and eastern regions have higher reddening as compared to western regions. We carried out cell-by-cell comparison of $E(V-I)$ obtained in our study with that of the \cite{2005AA...430..421S} as shown in Figure~\ref{comp_lmc}(a) with the dotted line which shows that reddening values estimated by \cite{2005AA...430..421S} is marginally underestimated in comparison of our studies.
\item \cite{2011AJ....141..158H} estimated reddening $E(V-I)$ using RR Lyrae stars and RC stars form OGLE-III data. They reported low reddening in central bar regions of the LMC and higher reddening towards 30 Doradus region s has also been noticed in the present study. In Figure~\ref{comp_lmc}(a), we show comparison of our estimated reddening values with those of \cite{2011AJ....141..158H} with continuous line and both the studies are found in close agreement.
\item \cite{2016MNRAS.463.1446N} studied 1072 star clusters in the LMC using OGLE-III data and a semi-automated quantitative method was used to estimate the age and reddening of these clusters. Their $E(V-I)$ values range from 0.05 to 0.50 mag with maximum reddening lying between 0.1 to 0.3 mag. The distribution of cell-by-cell comparison shown in Figure~\ref{comp_lmc}(a) with dashed line peaks at $\sim$-0.28 mag, which means that the reddening given by \cite{2016MNRAS.463.1446N} is much higher in comparison of any other studies carried out in the LMC in recent times.
\item \cite{2016ApJ...832..176I} investigated LMC disk using Cepheids in OGLE-IV data. In the Figure~\ref{comp_lmc}(b), we illustrate comparison between $E(B-V)$ values derived in present study and \cite{2016ApJ...832..176I}. The histogram shows that two studies are in excellent agreement within the quoted uncertainties. This result also endorse our approach of combining two modes of Cepheids in a single $P$-$L$ relation in order to estimate reddening measurements.
\end{enumerate}

%
%--------------- Figure 10 -------------------------
%--------------- End Figure -----------------------------%
%
In general, we find our reddening map is in excellent agreement with the recent reddening studies in the optical region except that of the \cite{2016MNRAS.463.1446N}. The histograms shown in Figure~\ref{comp_lmc} also provide opportunity to examine the difference in the reddening estimates traced through different stellar populations. It is seen from Figure~\ref{comp_lmc} that while \cite{2005AA...430..421S} and  \cite{2011AJ....141..158H} yield smaller reddening values from the older stellar populations RC and RR Lyrae stars, the map constructed by \cite{2016MNRAS.463.1446N} gives higher reddening values using the young star clusters in comparison of Cepheid variables which are relatively intermediate age stellar populations. It is also found that \cite{2016ApJ...832..176I}, who also used Cepheids in their study like in the present analysis, has similar reddening estimates as ours. In general, younger populations like open clusters are likely to be more embedded in the dusty clouds hence provide larger reddening than the older populations like RC stars and RR Lyraes.
\subsubsection{The SMC}\label{smc}
\begin{figure}[h]
\centering
\includegraphics[width=10.0cm,height=7.0cm]{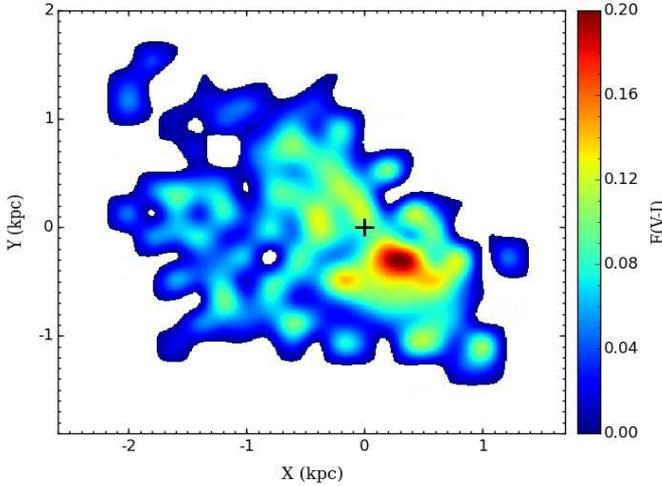}
\caption{Same as Figure~\ref{map_lmc} but for the SMC.}
\label{map_smc}
\end{figure}

The reddening values determined for 136 selected segments within the SMC was used to construct a reddening map which is illustrated in Figure~\ref{map_smc}. The optical center of SMC ($\alpha = 00^h 52^m 12^s.5 \equiv 13^{o}.05$ and $\delta = -72^{o} 49^{'} 43^{''} \equiv -72^{o}.82$; \cite{1972VA.....14..163D}) is shown in the  same figure. The reddening map exhibits a non-uniform and highly clumpy structure across the SMC. The peripheral regions of the SMC show smaller reddening in comparison of the central regions. The reddening is larger in the south west parts of the SMC. The largest reddening in the SMC is found to be $E(V-I) = 0.189$ in the region centred at $X$=0.3, $Y$=-0.3 which corresponds to $\alpha \sim 12^{o}.10,~\delta \sim -73^{o}.07$. On an average, the reddening in the SMC is found to be $E(V-I) = 0.049\pm0.070$ mag and corresponding $E(B-V) = 0.038\pm0.053$ mag which are substantially smaller in comparison of the mean reddening values in the LMC. If we compare the reddening maps in the LMC and SMC shown in Figures~\ref{map_lmc} and \ref{map_smc} with that of the spatial distributions of Cepheids given in Figures~\ref{spatial_lmc} and \ref{spatial_smc}, it is interesting to note that the reddening structures are found to be slightly correlated with those of the spatial distributions of Cepheids in both the galaxies, particularly in the SMC where higher reddening has been observed in the close vicinity of the densely populated regions of Cepheids.

%--------------- Figure 11 -------------------------
%--------------- End Figure ------------------------------
%--------------- Figure 12 -------------------------------
\begin{figure}[h]
\centering
\includegraphics[width=10.0cm,height=6.0cm]{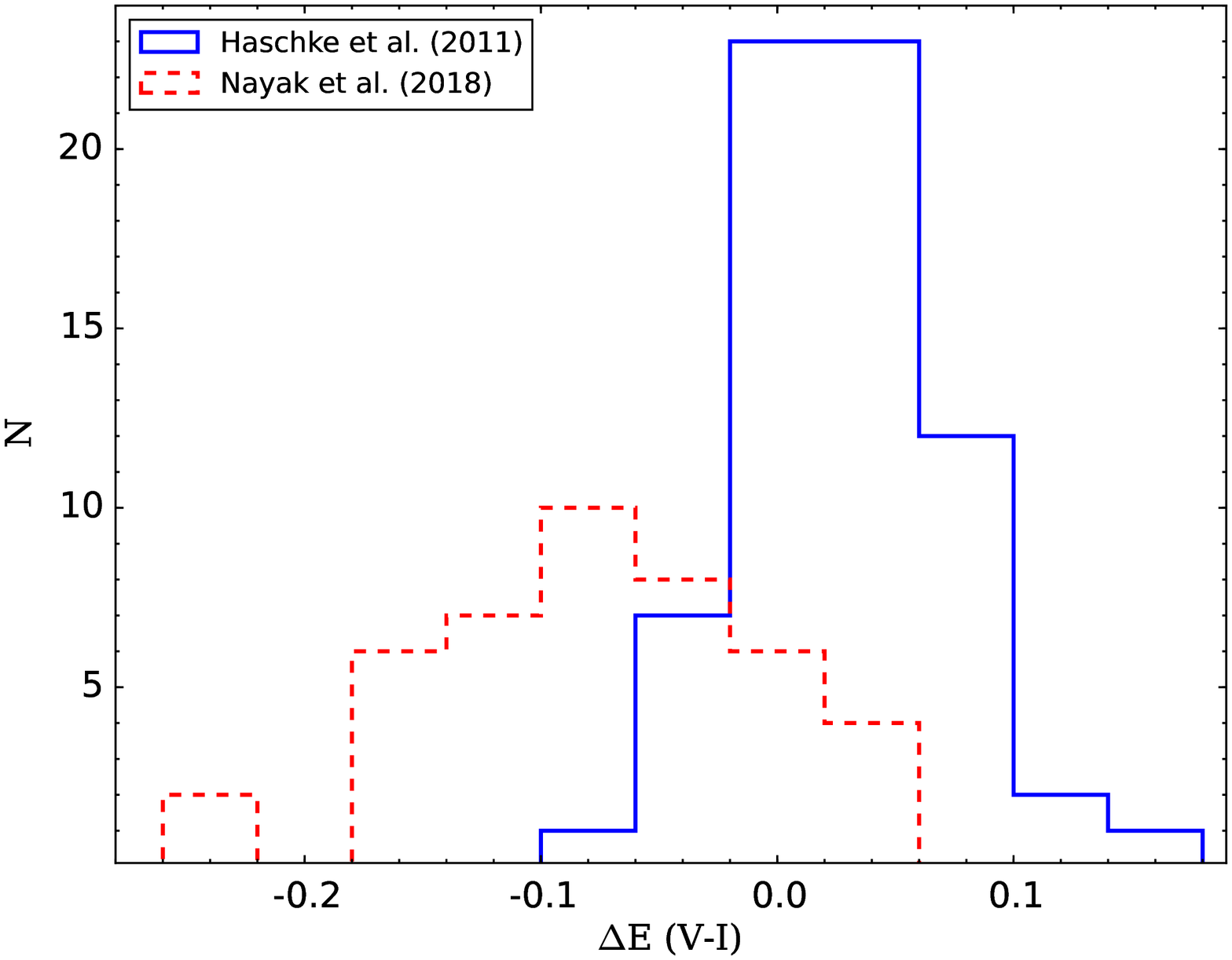}
\includegraphics[width=10.0cm,height=6.0cm]{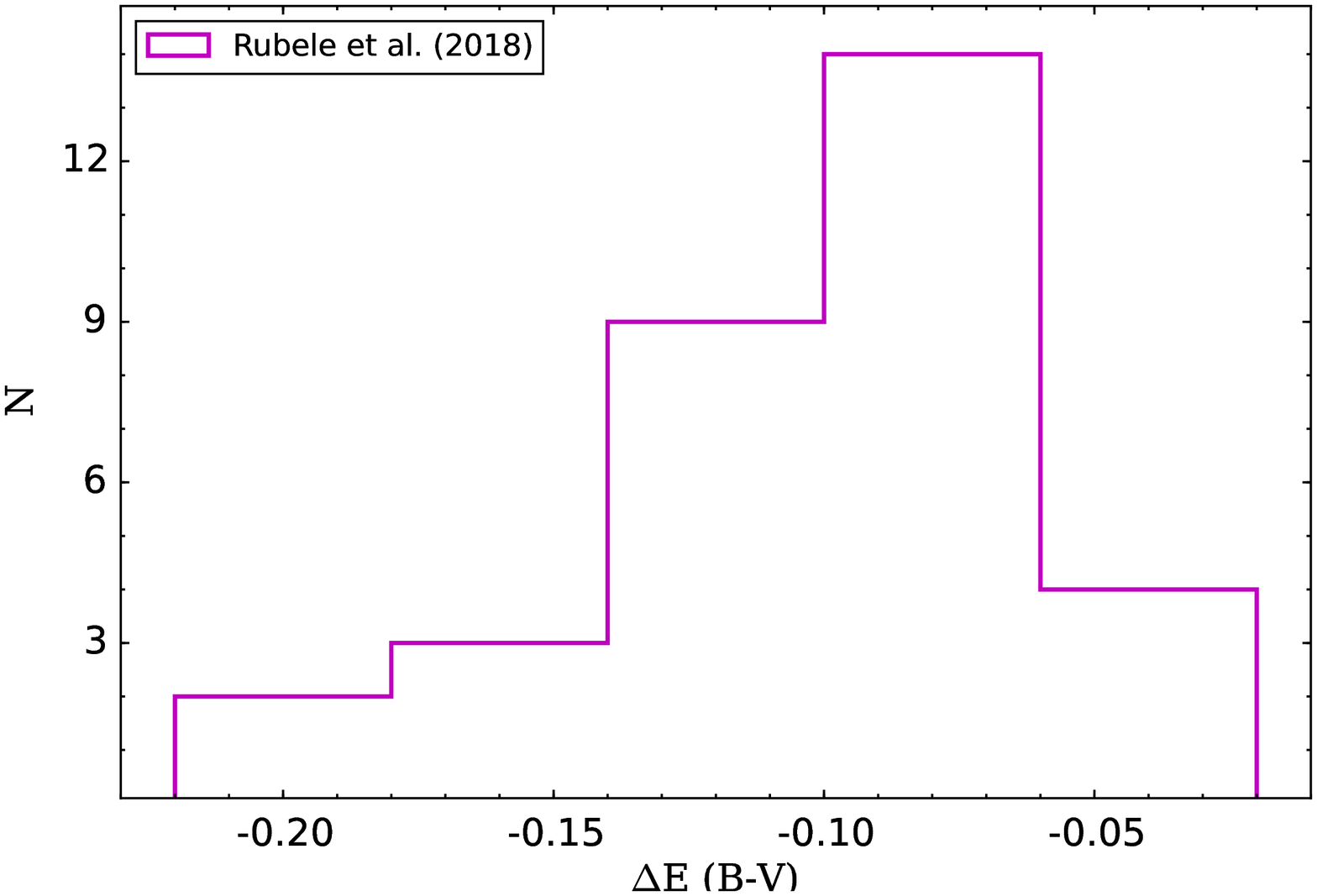}
\caption{(a) Upper panel shows the distributions of $\Delta E(V-I)$ i.e. past reddening estimates subtracted from our estimate in the SMC as mentioned on the top-left corner, (b) Lower panel shows the difference of $E(B-V)$ from \cite{2018MNRAS.478.5017R} subtracted from our reddening estimate.}

\label{comp_smc}
\end{figure}
%--------------- End Figure -------------------
%--------------- Figure 13 --------------------------
\begin{figure*}
\centering
\includegraphics[width=15.0cm,height=6.0cm]{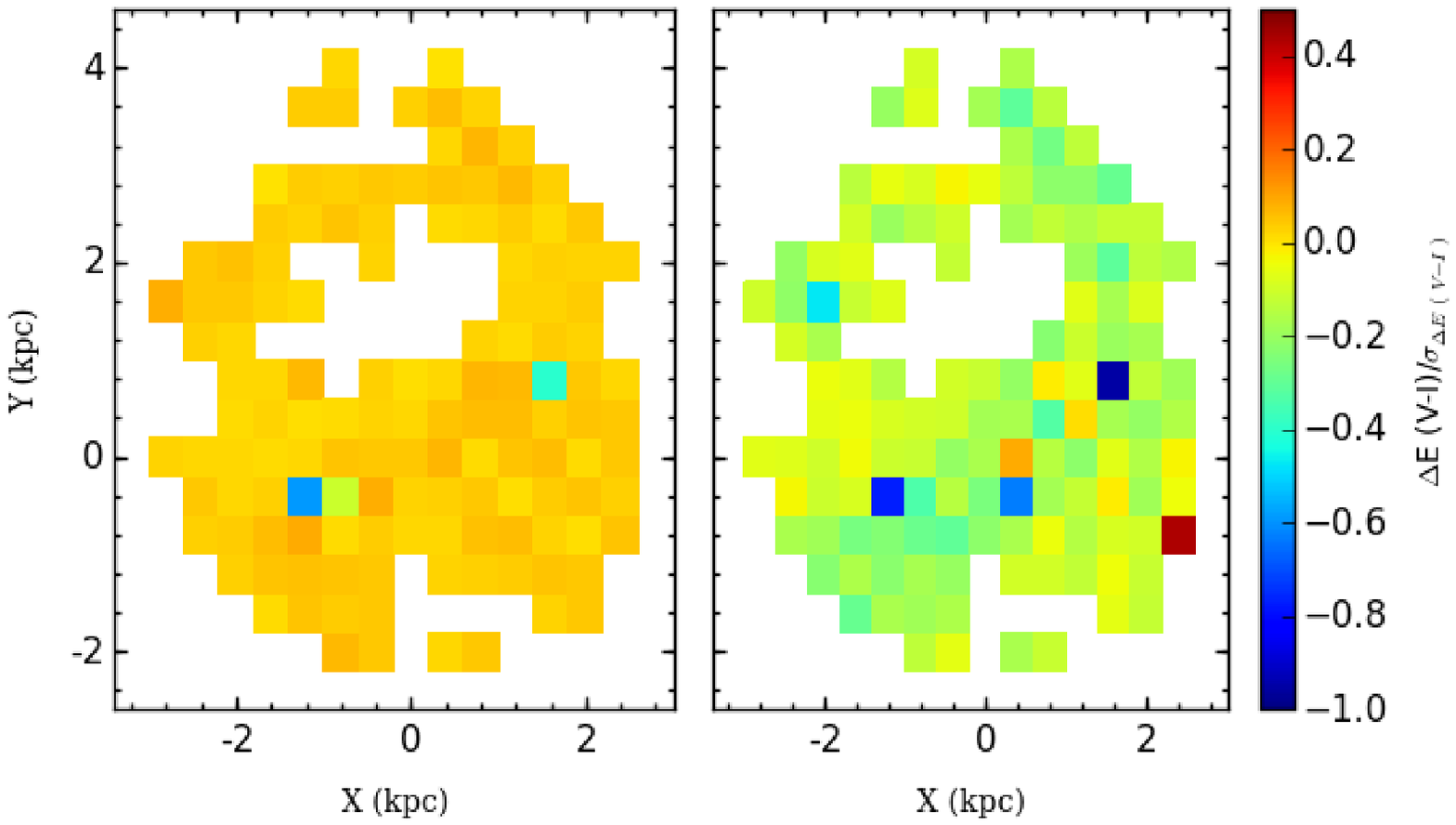}
\includegraphics[width=10.0cm,height=6.0cm]{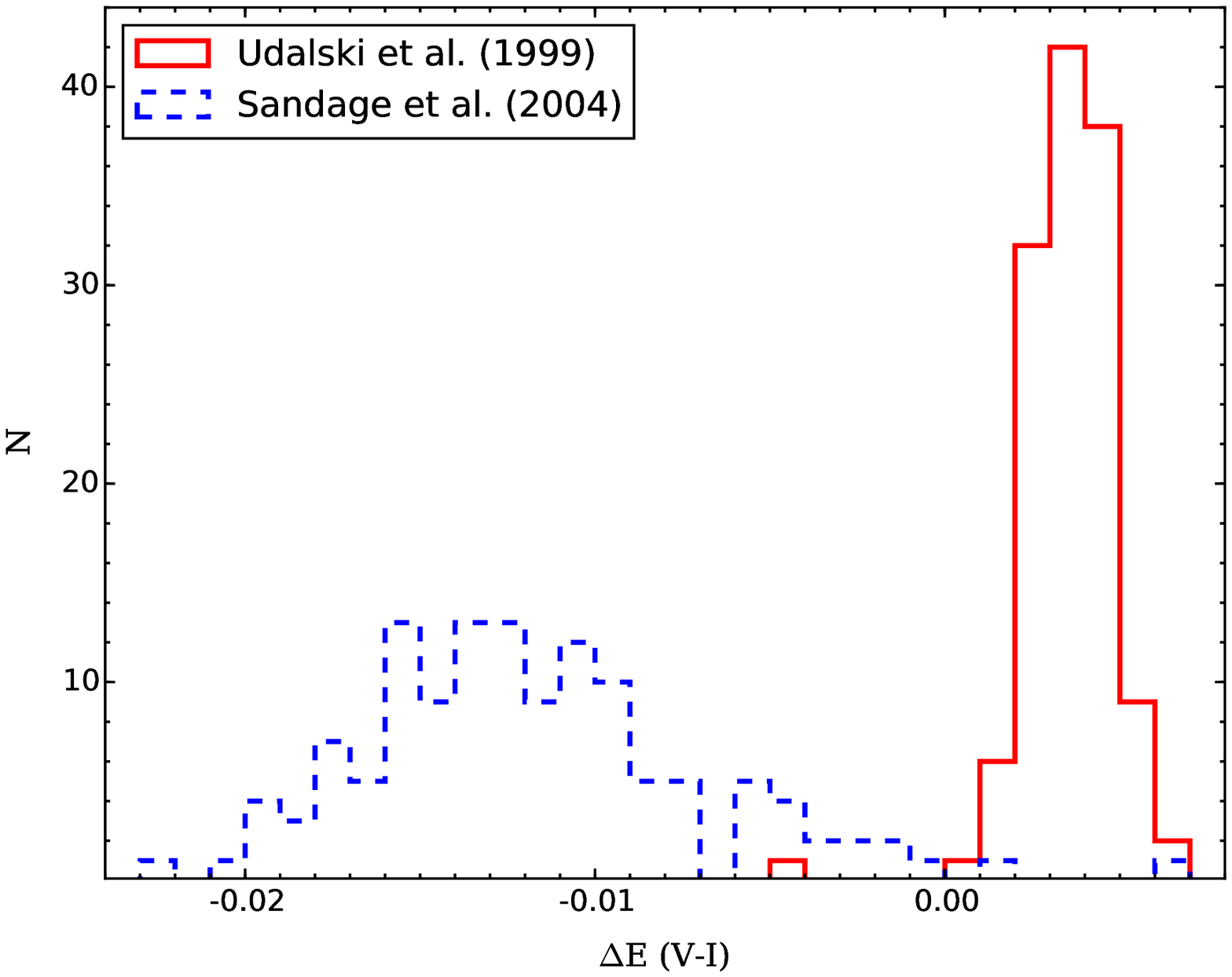}
\caption{Lower panel shows the histogram of the difference $\Delta~E(V-I)$ between the present study and those of \cite{1999AcA....49..201U} drawn by continuous line and \cite{2004A&A...424...43S} drawn by dashed line. Upper panel shows the heat map representing the difference in reddening divided by reddening uncertainty in each segment between the present reddening estimates and that of the  \cite{1999AcA....49..201U} in the left panel and same comparison between our values and  \cite{2004A&A...424...43S} in the right panel. The difference is obtained by subtracting other results out of our estimates.}
\label{comp_red_lmc}
\end{figure*}
%--------------- End Figure -----------------------------
%

The reddening maps of the SMC have been presented by many authors in the past using different kind of stellar populations \citep{1998ApJ...500..525S, 2002AJ....123..855Z, 2009AJ....137.5099D, 2011AJ....141..158H, 2015AA...573A.135S, 2018MNRAS.478.5017R, 2018A&A...616A.187N}. In the following part, we discuss a comparative study of our reddening map with some of the recent studies carried out in the SMC.
\begin{enumerate}
\item \cite{2011AJ....141..158H} studied the SMC reddening $E(V-I)$ using RR Layrae and RC stars from the OGLE-III data. The continuous line in Figure~\ref{comp_smc} shows difference between reddening $E(V-I)$ estimated by us and \cite{2011AJ....141..158H} using RC stars. The distribution peaks at 0.02 mag suggesting that \cite{2011AJ....141..158H} reddening values determined are slightly smaller in comparison of our study. 
\item \cite{2018A&A...616A.187N} studied 179 open clusters within the SMC using the same techniques as in \cite{2018A&A...616A.187N} carried out for the LMC open clusters. We show comparison of present reddening estimates with the \cite{2018A&A...616A.187N} in Figure~\ref{comp_smc} as a dashed line. The distribution peaks at -0.08 mag, which means that \cite{2018A&A...616A.187N} reddening values are larger in comparison of our reddening estimates. A similar study carried out earlier in the LMC has also shown that their reddening estimates are higher in comparison of the present estimates.
\item \cite{2018MNRAS.478.5017R} constructed the reddening map using 14 deep tile images taken in the $YJK_s$ filters under VMC survey and found a range of extinction ($A_V$) between $\sim$0.1 mag (external regions) and 0.9 mag (inner regions). On a comparative study between ours and \cite{2018MNRAS.478.5017R} reddening maps in the 32 common regions, we found that our $E(B-V)$ values are smaller in comparison of their study. In the lower panel of Figure~\ref{comp_smc}, we present histogram of the differences between two reddening estimates which shows a range of variations peaking around -0.08 mag. We however note that their study is based on near-IR data whereas our reddening estimates are based on the optical data.
\end{enumerate}

It is evident from the histograms in Figure~\ref{comp_smc} that \cite{2011AJ....141..158H} which used older stellar population found lower reddening values while \cite{2018A&A...616A.187N} which used younger stellar population retrieved higher reddening values in comparison of our reddening estimates in the SMC, a result similar to our previous comparison for the LMC. It is therefore quite clear from these comparative studies that the low reddening values follow the distribution of the older stellar populations and younger populations cater higher reddening in the MCs.
%

%--------------- Figure 14 --------------------------
\begin{figure*}
\centering
\includegraphics[width=15.0cm,height=6.0cm]{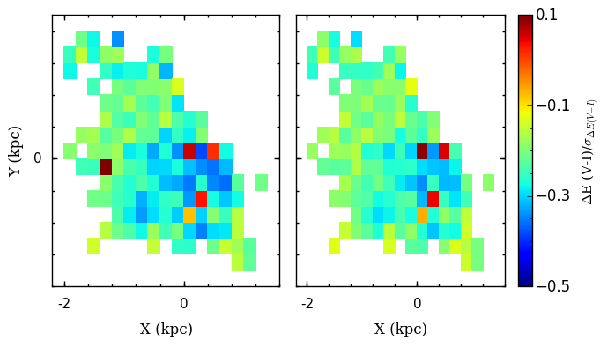}
\includegraphics[width=10.0cm,height=6.0cm]{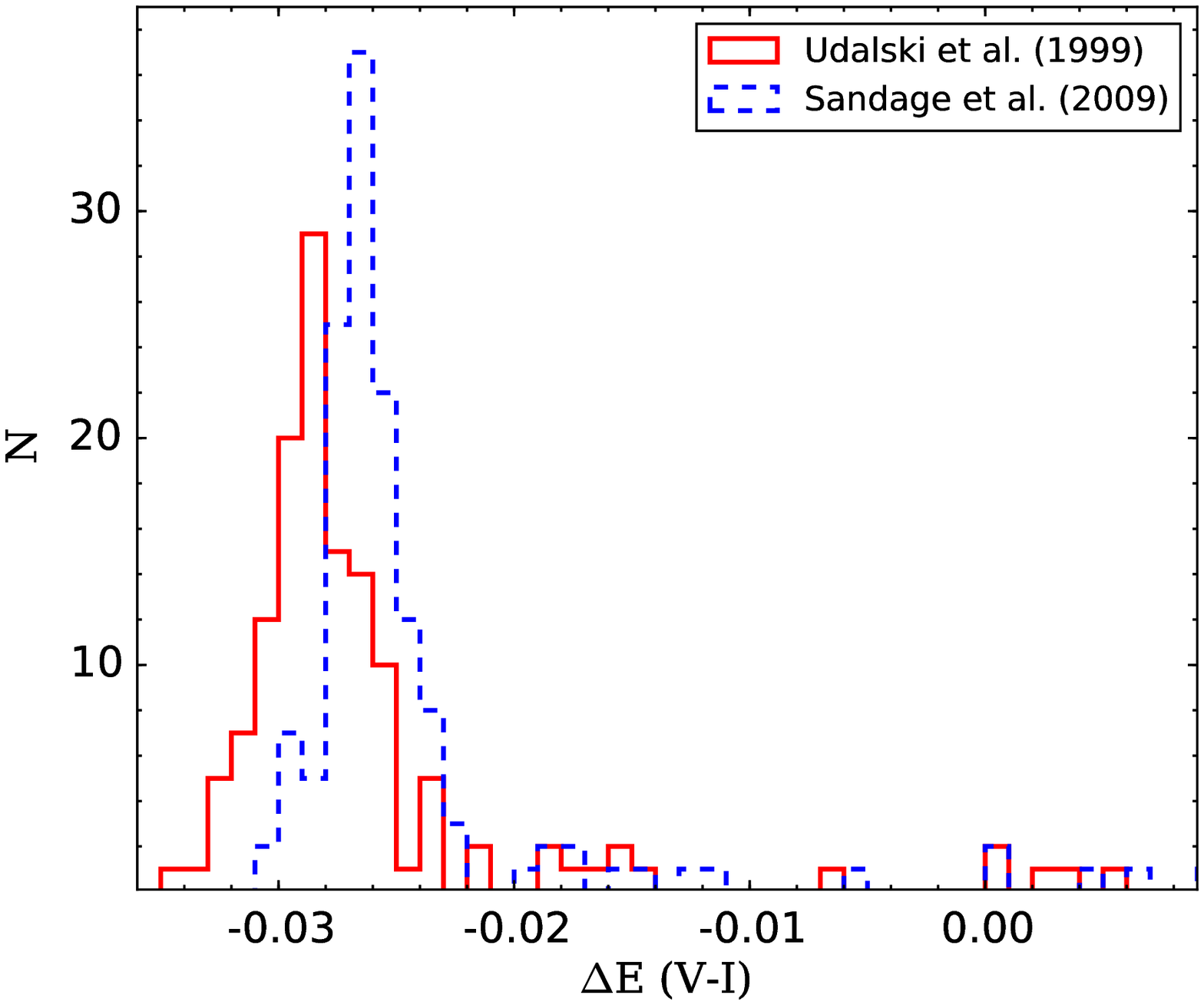}
\caption{Same as Figure~\ref{comp_red_lmc} but for the SMC. Here, comparisons are made between our maps and those obtained through \cite{1999AcA....49..201U} and \cite{2009A&A...493..471S} $P$-$L$ relations.}
\label{comp_red_smc}
\end{figure*}
%--------------- End Figure -----------------------------
%
\subsection{Effect on the Reddening Maps due to different $P$-$L$ relations}\label{effect}
In the present study, we used most recent $P$-$L$ relations given by \cite{2009ApJ...693..691N} and \cite{2015ApJ...808...67N} to determine reddening estimates in the regions of the LMC and SMC, respectively. There are many other $P$-$L$ relations reported in the previous studies such as \cite{1994MNRAS.266..441L}, \cite{1999AcA....49..201U}, \cite{2001ApJ...553...47F}, \cite{2004A&A...424...43S, 2009A&A...493..471S}, \cite{2009ApJ...693..691N, 2015ApJ...808...67N}. To investigate whether reddening distribution across the LMC and SMC determined in the present study changes due to different $P$-$L$ relations, we cross-examined present reddening maps with those using the \cite{1999AcA....49..201U} $P$-$L$ relations as well independent set of P-L relations given by \cite{2004A&A...424...43S} for LMC and \cite{2009A&A...493..471S} for the SMC. We used the same methodology as discussed previously in Section~\ref{method} to find reddening values in the MCs and constructed the reddening maps for both the LMC and SMC.

In the lower panel of Figure~\ref{comp_red_lmc}, we illustrate the histogram of the difference between present estimates with those determined through previous $P$-$L$ relations given by \cite{1999AcA....49..201U} and \cite{2004A&A...424...43S} by continuous and dashed lines, respectively. While we yield slightly lower values of $E(V-I)$ using \cite{1999AcA....49..201U}, we acquired slightly larger $E(V-I)$ values in comparison of \cite{2004A&A...424...43S, 2009A&A...493..471S}. It is seen that the reddening estimates in the LMC obtained through \cite{2009ApJ...693..691N} in the present study are in between to those obtained through \cite{1999AcA....49..201U} and \cite{2004A&A...424...43S} $P$-$L$ relations but this variation is less than 0.02 mag which is quite small in comparison of the uncertainties involved in the reddening estimations itself. For a quantitative verification of our reddening values, we also present significance of the difference among different reddening maps in the upper panel of Figures~\ref{comp_red_lmc} which illustrates the difference in reddening divided by the reddening uncertainty in each segment. Here, we determine comparison of our values with those obtained through the \cite{1999AcA....49..201U} and \cite{2004A&A...424...43S} $P$-$L$ relations in the left and right side, respectively. These maps suggest an excellent quantitative agreement between the reddening maps obtained through different $P$-$L$ relations in the LMC except few isolated segments as seen by blue squares in  Figure~\ref{comp_red_lmc}.

A similar comparison between the reddening values obtained in the present study using \cite{2015ApJ...808...67N} $P$-$L$ relations and those obtained through  \cite{1999AcA....49..201U} and \cite{2009A&A...493..471S} $P$-$L$ relations are illustrated in Figure~\ref{comp_red_smc} for the SMC. It is conspicuous from the differential plots that the present estimates in the SMC are slightly smaller than those obtained through both \cite{1999AcA....49..201U} and \cite{2009A&A...493..471S} $P$-$L$ relations however differences lie between 0.02 mag to 0.03 mag which can be considered as non-significant considering the uncertainties involved in the estimation of these values.  In general, we conclude that the change in Cepheids P-L relations does not make any noticeable difference in the reddening maps of these two clouds.
\section{Recent star formation history in the MCs}\label{distribution}
Recent studies shows that major star formation events took place in the MCs at several epochs ranging from few Gyrs to few Myrs ago \citep[e.g.,][]{2004AJ....127.1531H, 2014MNRAS.445.2214R, 2018MNRAS.478.5017R} though with varying star formation rates from field to field \citep{2013ApJ...775...83C, 2012A&A...537A.106R, 2015MNRAS.449..639R}. Therefore a comprehensive study of Cepheids provides a unique opportunity to probe the recent SFH in the MCs as these are relatively young population and most of them have ages less than few hundred Myr. Therefore the age and spatial-temporal distributions of Cepheids along with MCs structural parameters may provide important information about the formation history of the  Magellanic System.
\subsection{Age Distribution}\label{age_distri}
The $P$-$L$ and mass-luminosity relations of Cepheids imply that longer period Cepheids have higher luminosities and are more massive which means relatively shorter life span for these pulsating stars. Therefore, the period and age of Cepheids have an obvious connection. As Cepheids typically have ages in the range of roughly 30 to 600 Myr, a study of age distribution of Cepheids can be used to reconstruct the recent SFH within the MCs in last few tens to few hundreds Myr. Since pulsation period of Cepheids is the only quantity which can be precisely determined from the observations of the pulsating stars, their ages can be determined with a good accuracy using the PA relations.

To determine ages of Cepheids from their periods,\cite{1997A&AS..126..401M}, \cite{1998MNRAS.299..588E}, \cite{2003ARep...47.1000E} proposed many semi-empirical relations and \cite{2005ApJ...621..966B} provided theoretical period-age (PA) and and period-age-colour (PAC) relations. Recently \cite{2014NewA...28...27J} used mean periods of 74 LMC Cepheids found in 25 different open clusters and corresponding cluster ages taken from \cite{2010MNRAS.403.1491P} to draw an improved PA relation in the LMC. The empirical PA relation derived by \cite{2014NewA...28...27J} for the Cepheids in the LMC is given as
\begin{equation}
\log ({\rm t}) = 8.60(\pm0.07) - 0.77(\pm0.08)~\log ({\rm P})
\label{eqno10} 
\end{equation}
where age denoted by $t$ is in years and P is the period given in days. The reliability of the above PA relation has been examined in \cite{2014NewA...28...27J} by comparing this relation with the \cite{2005ApJ...621..966B} theoretical PA relation determined on the basis of evolutionary and pulsation models covering a broad range of stellar masses and chemical compositions. We found a reasonable agreement between the two relations given for the LMC.

Cepheids PA relation however varies for the galaxies having difference metallicities and as we have not derived any such PA relation for the SMC, we considered here theoretical PA relation given by the  \cite{2005ApJ...621..966B} for the SMC. Since we have already converted period of FO Cepheids to corresponding FU Cepheids, we only used PA relation of FU Cepheids for the known metallicity of SMC ($z$=0.004):
\begin{equation}
\log ({\rm t}) = 8.49(\pm0.09) - 0.79(\pm0.01)~\log ({\rm P})
\label{eqno11} 
\end{equation}

We used above PA relations to determine the age of each Cepheid in the LMC and SMC. As error in the period of Cepheids is reported to be less than 0.001\%, we have not taken them into account in the subsequent age conversion but the typical average error in age is estimated to be $\sim$40 Myr due to uncertainty in the PA relations given by \cite{2005ApJ...621..966B} and \cite{2014NewA...28...27J}. Although resulting ages are in the range of $\log ({\rm t/yr})$ = 6.96 to 8.96 with a mean $\log ({\rm t/yr})$ of 8.21 for the LMC but three-fourth Cepheids population in the LMC are distributed between $\log ({\rm t/yr})$ of 8.0 to 8.4.  Similarly, ages for the SMC Cepheids ranges from $\log ({\rm t/yr})$ = 6.66 to 8.86 with a mean age of  8.27, about three-fourth of them are confined between $\log ({\rm t/yr})$ of 8.1 to 8.5. One can see that the SMC Cepheids are on average slightly older in comparison of the LMC  Cepheids.
 
\begin{figure}
\centering
\includegraphics[width=10.0cm,height=6.5cm]{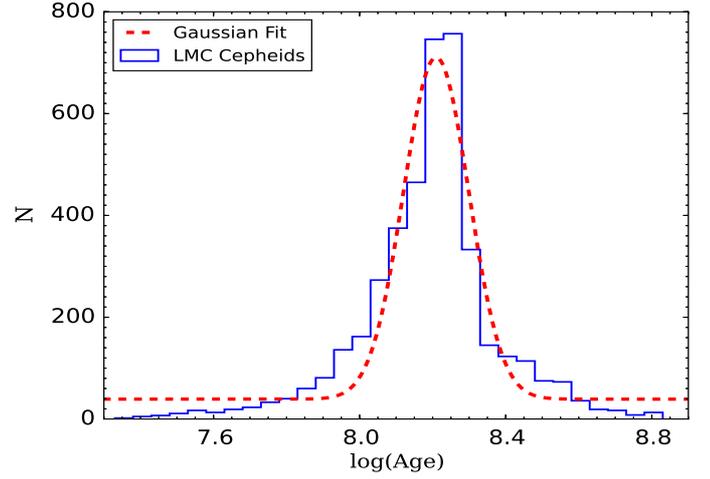}
\caption{The age distribution of Cepheids in the LMC. The best fit Gaussian profile is shown by a dashed red line.}
\label{ceph_age_lmc}
\end{figure}

\begin{figure}
\centering
\includegraphics[width=10.0cm,height=6.5cm]{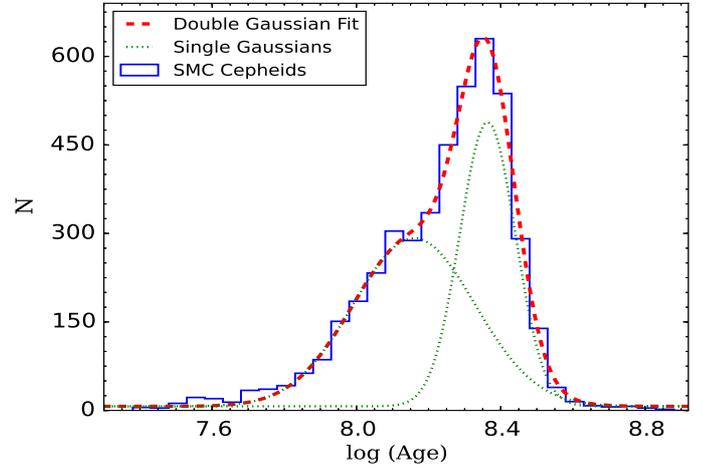}
\caption{Same as Figure~\ref{ceph_age_lmc} but for the SMC. The best fit of double Gaussian is shown as a dashed red line and individual Gaussian are shown by dotted green line.}
\label{ceph_age_smc}
\end{figure}

We determined the distribution of Cepheids in a bin width of 0.05 dex (on a logarithmic scale) in the LMC and SMC which are respectively shown in Figures~\ref{ceph_age_lmc} and \ref{ceph_age_smc}. We see a pronounced peak in the age distribution of Cepheids in the LMC that can be represented by a Gaussian-like profile. In Figure~\ref{ceph_age_lmc}, we show the best fit Gaussian distribution in the histogram that gives a peak at $\log ({\rm t/yr}) = 8.21\pm0.11$. However, a slightly broad profile is evident in the age distribution of Cepheids in the SMC which may be represented by a bi-modal Gaussian profile as has been drawn by \cite{2016AcA....66..149J}. Though we do not observe a clear bimodality in the age distribution of SMC Cepheids but a double-Gaussian profile still gives a better fit than a single-Gaussian therefore we prefer the earlier one. A best fit double-Gaussian profile in the age histogram, which is shown by a dashed line in Figure~\ref{ceph_age_smc}, gives a prominent primary peak at $\log ({\rm t/yr}) = 8.36\pm0.08$ and a small secondary peak at $\log ({\rm t/yr}) = 8.17\pm0.08$. The peak for the younger Cepheids is not strong enough to draw any firm conclusion although it is coinciding with the primary peak in the LMC. We note here that the final error in age represents the combined uncertainty in the PA relation as well as error corresponds to the mean age estimation in the Gaussian fit. The maxima in age distributions of Cepheids indicate a rapid enhancement of Cepheid formation at around $162_{-36}^{+46}$ Myr for the LMC while it is $229_{-39}^{+46}$ Myr for the SMC. The rapid enhancement of Cepheids around 200 Myr ago (within 1-$\sigma$ uncertainty) in the LMC and SMC thus pointed to a major star formation episode at that time, most likely due to a close encounter between the two clouds or their interactions with the Galaxy stem from their multiple pericentric passages as they orbit the Milky Way.

%
%---------------------------- Fig. 15---------------------------------------
 %---------------------------- End ----------------------------
%---------------------------- Fig. 16---------------------------------------
 %---------------------------- End ----------------------------

The oscillation between rise and fall in the star formation rate depends upon whether these two clouds are approaching or receding \citep{2010A&A...517A..50G, 2016RAA....16...61J}. This repeated interaction between LMC and gas-rich SMC lead to episodic star formations in these two dwarf galaxies which are locked in tidal interaction. It is believed that the Magellanic Bridge and the Magellanic Stream might have formed due to such interactions in the past between the two segments of the MCs \citep{2010ApJ...721L..97B, 2018MNRAS.473.3131M}. As a consequence of these frequent encounters, the tidal stripping of stars and gas/material from the gaseous disk takes place in the Magellanic System. In fact gas in the Magellanic Bridge is thought to have been largely stripped from the SMC as a consequence of its close interactions with the LMC at about 200 Myr ago \citep{2014MNRAS.442.1680D, 2017MNRAS.472.2975M}. 
 
Considering several previous studies on the interaction of the clouds \citep[e.g.,][]{2012MNRAS.421.2109B, 2018ApJ...864...55Z, 2019ApJ...874...78Z} there are ample evidences of a direct collision between two clouds with an impact parameter of few kpc \citep{2018ApJ...867L...8O, 2018ApJ...864...55Z}. These studies along with many previous studies confirm epochs of recent star formation in the MCs albeit at slightly different ages. Using open star clusters in the SMC, \cite{2000AcA....50..337P} and \cite{2010A&A...517A..50G} found a peak at 160 Myr through their age distribution. A similar conclusion was also drawn by \cite{2015AA...573A.135S} using OGLE-III data based on \cite{2005ApJ...621..966B} PAC relation and \cite{2016AcA....66..149J} using OGLE-IV data based on \cite{2005ApJ...621..966B} PA relation. In a recent investigation by \cite{2017MNRAS.472..808R} using the data from VISTA near-infrared YJKs survey of the Magellanic System (VMC), they also predicted a close encounter or a direct collision between the two cloud components some 200 Myr ago and confirm the presence of a Counter-Bridge. 

On the theoretical side, there have been several studies to infer cloud-cloud interaction in the MCs. In some of the recent models proposed by \cite{2005MNRAS.356..680B}, \cite{2006ApJ...652.1213K}, \cite{2012ApJ...750...36D}, \cite{2012MNRAS.421.2109B} and \cite{2018ApJ...864...55Z}, they suggested the last cloud-cloud collision within the Magellanic System had occurred about 100-300\,Myr ago causing an enhanced star formation activities in these two dwarf galaxies that might have resulted the formation of Magellanic Stream. In fact  mutual interactions between the clouds and subsequent tidal stripping of material from the SMC are believed to be the most likely reason for the formation of Magellanic Bridge \citep{2012MNRAS.421.2109B, 2012ApJ...750...36D, 2018MNRAS.473.3131M}. This would also mean that Magellanic stream and Magellanic bridge stellar populations should contain stars from both the components of the MCs. On the other hand, it was also suggested by \cite{2010A&A...517A..50G} (and references therein) that not only the frequent tidal interaction between the two clouds but stellar winds and supernova explosions may also induce episodic star formation in these two dwarf galaxies. 
\subsection{Spatio-temporal Distribution}\label{spatial_tempo_distri}
\begin{figure}[h]
\centering
\includegraphics[width=10.0cm,height=7.0cm]{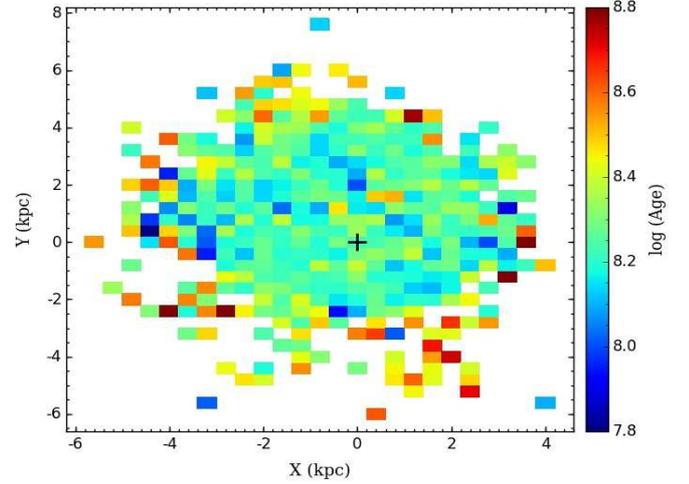}
\caption{Spatial distributions of the Cepheids in the 389 segments of the LMC as a function of their mean $\log ({\rm t/yr})$. North is up and east is to the left. The locations of the optical center of LMC is shown by plus sign.}
\label{spatial_age_lmc}
\end{figure}

\begin{figure}[h]
\centering
\includegraphics[width=10.0cm,height=7.0cm]{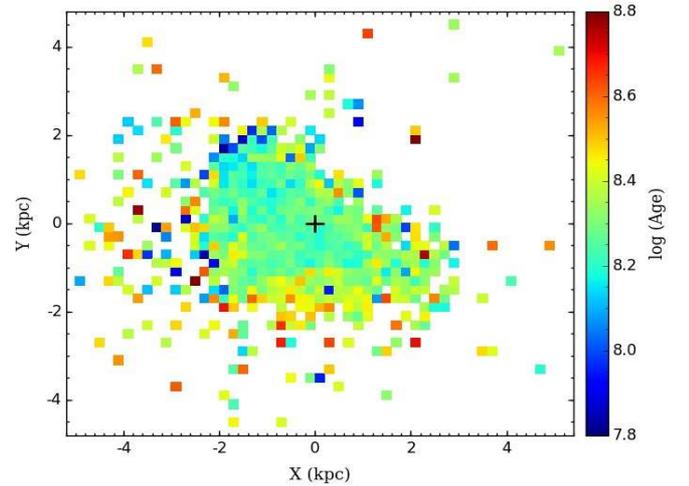}
\caption{Same as Figure~\ref{spatial_age_lmc} but for the 562 segments in the SMC.}
\label{spatial_age_smc}
\end{figure}

The spatial distribution of Cepheids as a function of age is shown in Figures~\ref{spatial_age_lmc} and \ref{spatial_age_smc} for the LMC and SMC, respectively. The figures show that the Cepheids are distributed all over the LMC while preferential distributions of Cepheids is seen in the SMC. From the examination of age map of the LMC, complex and patchy nature in age distribution is quite evident where both young and old Cepheid populations are distributed in small structures across the cloud. We also see a gradual change in the ages of Cepheids from central to peripheral regions where inner regions have lower ages and outer regions have pockets of higher ages as has been noticed in the last star formation event (LSFE) map presented by the \cite{2011A&A...535A.115I} through the determination of the Main-Sequence (MS) turn-off point in the colour-magnitude diagram. A similar structure has also been observed by \cite{2014MNRAS.438.1067M} who suggested a stellar population gradient in the LMC disk where younger stellar populations are more centrally concentrated. They also proposed an outside-in quenching of the star formation in the outer LMC disk ($R_{GC}$ = 3.5-6.2 kpc) which might be associated with the variation of the size of HI disk as a result from gas depletion due to star formation or ram-pressure stripping, or from the compression of the gas disc as ram pressure from the Milky Way halo acted on the LMC interstellar medium.

%--------------- Figure 17--------------------------
%--------------- End Figure -----------------------------

In the SMC, age map of Cepheids shows a systematic distribution where younger Cepheids lie towards inner region and older Cepheids are mainly confined towards peripheral regions. This suggest an inwards quenching of star formation in the SMC. \cite{2017MNRAS.472..808R} also found that young and old Cepheids have different geometric distributions in the SMC. They observed that closer Cepheids are preferentially distributed in the eastern regions of the SMC which are off-centred in the direction of LMC owing to the tidal interaction between the two clouds \citep{2018MNRAS.473.3131M}.

%--------------- Figure 18--------------------------
%--------------- End Figure -----------------------------

If we compare frequency distribution map of the Cepheids as shown in Figures~\ref{spatial_age_lmc} and \ref{spatial_age_smc} with the similar maps made for the star clusters identified in \cite{2010A&A...517A..50G} (see Figure~9 of \cite{2014NewA...28...27J} for the LMC, and Figure~5 of \cite{2016RAA....16...61J} for the SMC), we notice that the clumps of Cepheids do not coincide with the clumps of the star clusters in both the component of the MCs and a mutual avoidance of clumps of the Cepheids and star clusters is present within the MCs. 
\section{Discussion and Conclusions}\label{conclu}
The nearby LMC and SMC are the two galaxies for which the highest number of Cepheids are detected, that too with an excellent data quality. The main motive of this paper was thus to exploit the available $V$ and $I$ band data of close to nine thousand Cepheids in the MCs. The independent reddening values determined through the multi-wavelength $P$-$L$ relations of Cepheids are important to construct reddening map which is necessary to understand the dust distributions within different regions of the host galaxy. In the present study $V$ and $I$ band photometric data of Cepheids provided in the OGLE-IV photometric survey was analysed to understand the reddening distributions across the LMC and SMC and subsequently to draw reddening maps in these two nearby galaxies. We used 2476 FU and 1775 FO Cepheids in the LMC and 2753 FU and 1793 FO Cepheids in the SMC in the present analysis. In the present study there is one major addition that instead of studying individual $P$-$L$ relations for FU and FO Cepheids, we combined these two modes of pulsating stars, after converting periods of FO Cepheids to the corresponding periods of FU Cepheids. This has increased our sample size of Cepheids in order to draw $P$-$L$ diagrams that has allowed us to make small size segments within the galaxies hence better resolution of the reddening maps. To reduce statistical error in the reddening estimation, we selected only those segments which contain a minimum of 10 Cepheids. We drew a best fit $P$-$L$ relation in both $V$ and $I$ band data in all the segments of the LMC as well as those of the SMC. Using the well calibrated $P$-$L$ relations for the LMC and SMC, we estimated the reddening $E(V-I)$ in each segment of both the galaxies. We found that the reddening $E(V-I)$ varies from 0.041 mag to 0.466 mag in the LMC and 0.00 to 0.189 mag in the SMC. The mean value of reddening obtained through best fit log-normal profile was found to be $E(V-I)=0.113\pm0.060$ mag and $E(B-V)=0.091\pm0.050$ mag for the LMC. The mean value of reddening obtained through similar approach in the SMC was estimated to be $E(V-I)=0.049\pm0.070$ mag and $E(B-V)=0.038\pm0.053$ mag.

Using the reddening distributions of 133 segments in the LMC and 136 segments in the SMC, we constructed reddening maps with a cell size of 0.4$\times$0.4 square kpc in the LMC and 0.2$\times$0.2 square kpc in the SMC. This provides an average angular resolution of about 1.2 deg$^2$ covering an area over 150 deg$^2$ in the LMC and angular resolution of about 0.22 deg$^2$ covering an area of over 30 deg$^2$ in the SMC. The reddening map of the LMC shows a heterogeneous distribution having small reddening in the bar region and higher reddening towards north-east region. We did not find any significant reddening in the central region of the LMC. The highest reddening in the LMC having $E(V-I)=0.466$ mag was traced in the north east region located at $\alpha \sim 85^{o}.13,~\delta \sim -69^{o}.34$ which seems to be associated with the most active star forming HII region 30 Doradus situated at $\alpha \sim 84^o$, $\delta \sim -70^o$ that also contains the highest concentration of Cepheids in our sample. In case of the SMC, reddening map exhibits a non-uniform and highly clumpy structure across the cloud. In general, smaller reddening was found around central regions of the SMC but larger reddening was seen in the south-west region away from the optical center. The largest reddening in the SMC is found to be $E(V-I) = 0.189$ in the region centred at $\alpha \sim 12^{o}.10,~\delta \sim -73^{o}.07$. The peripheral regions of the SMC have shown very small reddening. If we compare our reddening maps with those of the spatial maps in the MCs, we found a broad correlation between the denser regions to the reddened structures which is found to be more closely related in the SMC where higher reddening has been found in the close vicinity of the densely populated regions of Cepheids. The comparison of our reddening maps with the some recent optical reddening maps has also been carried out, most of them matches well with our results although few of them are found to be overestimated or underestimated in comparison of the present study. There are various stellar populations which are used to define the reddening map but shows some discrepancies among different populations because different stellar populations experience different dust. It is well expected that the reddening estimates through early type stars or star forming regions yield higher reddening in comparison of intermediate or old age stellar populations. Different stellar populations have different spatial distributions and most of them are non-axisymmetric. Furthermore, no significant variation was noticed in the reddening maps when we used different sets of $P$-$L$ relations which demonstrate the stability of present reddening maps.

The unprecedented large data set on Cepheids in the OGLE-IV survey led to refine our knowledge about the spatial and age distributions of Cepheids within the MCs. We found that Cepheids in the LMC and SMC are concentrated in the south-east and south-west regions, respectively. While dense population of Cepheids lies in close vicinity of the optical center of the SMC, it is substantially shifted from the optical center of the LMC. The western region of the LMC bar is densely populated for Cepheids in comparison of the eastern region and northern arm of the LMC reveals a very poor spatial density. To explore the recent star formation activity across the MCs, we also estimated the ages of Cepheids taking advantage of known PA relations in the literature. The age distribution in the LMC shows a Gaussian profile having peak at $\log ({\rm t/yr}) = 8.21\pm0.11$, the age distribution in the SMC displays a prominent peak at $\log ({\rm t/yr}) = 8.36\pm0.08$ although it shows a weak bimodal distribution. The age maxima in the LMC is found to be very close to that of the SMC which suggests a common enhancement of star formation had happened in these two galaxies sometime around 200 Myr ago. The most likely scenario for this simultaneous star formation burst is thought to be resulted due to cloud-cloud encounter between the LMC and SMC. A similar distribution has also been noticed in earlier studies using different stellar populations and our result also supports the current theoretical scenario predicting a close encounter between the Clouds. Although cloud-cloud collisions are well expected between these two nearby galaxies due to their tidal interactions but any external phenomenons like stellar winds and supernova explosions cannot be ruled out as a cause of enhanced star formations in these systems. We noticed a slightly preferential distribution in the SMC where relatively older Cepheids were observed towards the peripheral regions. It was interesting to note that eastern part of the SMC possessed most of younger Cepheids which indicates that the eastern region of the galaxy may be relatively younger.

As MCs have shown evidence of undergoing numerous star formation episodes in the past ranging from few Myrs to few Gyrs, it is absolutely necessary to study these two nearby clouds with different stellar populations as life span of these tracers also varies from few Myrs to few Gyrs. Moreover the properties of Clouds vary both in spatial distributions as well as a function of stellar population. Once derived, they can provide important clues to understand the outlying mechanisms of galaxy interactions that in turn drive the star formations over varying time scales. Therefore, in order to retrieve a complete picture of the Magellanic System that comprises LMC, SMC, Magellanic Bridge and Magellanic Stream, a combination of various data samples and multi-band catalogues of different stellar populations, particularly extracted through large sky surveys, is of vital importance. 
\section*{Acknowledgments}
We are grateful to the referee for providing helpful comments that significantly improved this paper. We are thankful to P. K. Nayak for providing their reddening estimates in the SMC before publication, Jeewan C. Pandey for his useful suggestions and Smitha Subramanian for pointing out few mistakes in the early stage of the draft. We also acknowledge Aurobinda Ghosh who repeated part of the present work during his summer project sponsored by the Indian Academy of Sciences (IASc), Bangalore through the grant no. IAS-SRFP 2018. This publication makes use of data products from the OGLE archive.
\bibliography{joshi}
\end{document}